\documentclass[
  journal=largetwo,
  manuscript=research-article,
  year=2025,
  volume=00,
]{cup-journal}

\usepackage{aas-macros}
\usepackage{amsmath,amssymb}
\usepackage[nopatch]{microtype}
\usepackage{booktabs}
\usepackage{hyperref}
\usepackage[range-phrase = \text{--}]{siunitx}
\usepackage{caption}
\usepackage{subcaption}

\usepackage[australian]{babel}
\usepackage{csquotes}
\hypersetup{colorlinks, citecolor=blue, linkcolor=blue, urlcolor=blue}

\DeclareSIUnit{\yr}{yr} 
\DeclareSIUnit{\h}{h} 
\DeclareSIUnit{\pc}{pc} 
\DeclareSIUnit{\TB}{TB} 
\DeclareSIUnit{\PB}{PB} 

\newcommand{\snr}[0]{$\mathrm{S}/\mathrm{N}$}

\title{The Southern-sky MWA Rapid Two-metre (SMART) pulsar survey---III. A~census of millisecond pulsars at 154\,MHz}

\author{C.~P.~Lee} 
\affiliation{International Centre for Radio Astronomy Research, Curtin University, Bentley, WA 6102, Australia}
\email[C.~P.~Lee]{christopher.lee@icrar.org}

\author{N.~D.~R.~Bhat} 
\affiliation{International Centre for Radio Astronomy Research, Curtin University, Bentley, WA 6102, Australia}

\author{B.~W.~Meyers} 
\affiliation{International Centre for Radio Astronomy Research, Curtin University, Bentley, WA 6102, Australia}
\alsoaffiliation{Australian SKA Regional Centre (AusSRC), Curtin University, Bentley, WA 6102, Australia}

\author{S.~J.~McSweeney} 
\affiliation{International Centre for Radio Astronomy Research, Curtin University, Bentley, WA 6102, Australia}

\author{W.~van~Straten} 
\affiliation{Manly Astrophysics, 15/41-42 East Esplanade, Manly, NSW 2095, Australia}

\author{C.~M.~Tan} 
\affiliation{International Centre for Radio Astronomy Research, Curtin University, Bentley, WA 6102, Australia}

\author{M.~Xue} 
\affiliation{National Astronomical Observatories, Chinese Academy of Sciences, 20A Datun Road, Chaoyang District, Beijing 100101, People's Republic of China}

\author{N.~A.~Swainston} 
\affiliation{International Centre for Radio Astronomy Research, Curtin University, Bentley, WA 6102, Australia}
\alsoaffiliation{CSIRO Space and Astronomy, PO Box 1130, Bentley, WA 6102, Australia}

\author{S.~M.~Ord} 
\affiliation{CSIRO Astronomy and Space Science, PO Box 76, Epping, NSW 1710, Australia}

\author{G.~Sleap} 
\affiliation{International Centre for Radio Astronomy Research, Curtin University, Bentley, WA 6102, Australia}

\author{S.~E.~Tremblay} 
\affiliation{National Radio Astronomy Observatory, 1011 Lopezville Road, Socorro, NM 87801, USA}

\author{A.~Williams} 
\affiliation{International Centre for Radio Astronomy Research, Curtin University, Bentley, WA 6102, Australia}

\keywords{instrumentation: interferometers -- methods: observational -- techniques: polarimetric -- pulsars: general} 

\begin{document}
    \begin{abstract}
Observations of millisecond pulsars (MSPs) at low radio frequencies play an important role in understanding the Galactic pulsar population and characterising both their emission properties and the effects of the ionised interstellar medium on the received signals.
To date, only a relatively small fraction of the known MSP population has been detected at frequencies below \qty{300}{\MHz}, and nearly all previous MSP studies at these frequencies have been conducted with northern telescopes.
We present a census of MSPs in the SMART pulsar survey, covering declinations south of $+\qty{30}{\degree}$ at a centre frequency of \qty{154}{\MHz}.
We detected 40 MSPs, with 11 being the first published detections below \qty{300}{\MHz}.
For each detection, we provide coherently-dedispersed full-polarimetric integrated pulse profiles and mean flux densities.
We measured significant Faraday rotation measures (RMs) for 25 MSPs, and identified apparent phase-dependent RM variations for three MSPs.
Comparison with published profiles at other frequencies supports previous studies suggesting that the pulse component separations of MSPs vary negligibly over a wide frequency range due to their compact magnetospheres.
We observe that integrated pulse profiles tend to be more polarised at low frequencies, consistent with depolarisation due to superposed orthogonal polarisation modes.
The results of this census will be a valuable resource for planning future MSP monitoring projects at low frequencies, and will also help to improve survey simulations to forecast the detectable MSP population with the SKA-Low.
\end{abstract}
    \section{Introduction}
Millisecond pulsars (MSPs) are a subgroup of rotation-powered pulsars with short spin periods ($P\sim\numrange{1}{10}\,\unit{\ms}$) and low spin-down rates ($\dot{P}\sim\numrange{e-4}{e-2}\,\unit{\ns\per\yr}$), placing them in a region of the $P$--$\dot{P}$ parameter space separate to normal pulsars ($P\sim\numrange{0.1}{10}\,\unit{\s}$, $\dot{P}\sim\numrange{e-1}{e3}\,\unit{\ns\per\yr}$).
MSPs are formed via a recycling process, in which an old slowly-rotating neutron star is spun up by mass transfer from a binary companion \citep{Alpar1982,Rad1982,Bhattacharya1991}.
As a result, almost all MSPs are found in binary systems.
Due to the high intrinsic stability of their average phase-resolved emission, MSPs are extremely precise astrophysical clocks which have several high-profile applications.
Notably, timing of MSPs in relativistic binaries can be used to probe the limits of strong-field gravity with unparalleled precision \citep[e.g.][]{Kramer2021MNRAS,Kramer2021PhRvX}, and regular observations of multiple MSPs as part of a pulsar timing array can be used to detect and study \unit{\nano\Hz}-frequency gravitational waves, including the stochastic gravitational-wave background \citep[e.g.][]{PPTA_GWB_2023,NANOGrav_GWB_2023,EPTA_GWB_2023}.
Characterising the observational properties of MSPs and the effects of the ionised interstellar medium (IISM) on the received signals is critically important for pulsar timing, and can help to increase the sensitivity of pulsar timing arrays \citep[e.g.][]{Cordes2010}.

The differences in radio emission properties between non-recycled pulsars and MSPs have been investigated in detail by several authors.
\citet{Kramer1998,Xilouris1998} found that MSPs are less luminous, have flatter position angle curves, and marginally more complex pulse profiles that exhibit less evolution with frequency.
Using the large data set provided by the MeerTime project, \citet{Karastergiou2024} found that the differences in radio luminosity are not significant when scaled to account for the pulse duty cycle and the solid angle of the radio beam.
Both \citet{Kramer1998} and \citet{Karastergiou2024} concluded that the flux density spectra of MSPs and non-recycled pulsars are similarly steep.
However, the spectra of non-recycled pulsars often flatten or turn over at around \numrange{100}{300}\,\unit{\MHz} \citep[e.g.][]{Sieber1973,Jankowski2018,Lee2022}, whereas very few MSPs show convincing evidence of turning over, and those that do are estimated to turn over below \qty{100}{\MHz} \citep{Kuzmin2001,Dowell2013,Kuniyoshi2015}.
\citet{Kramer1999} suggested that the lack of complexity in the spectra of MSPs is indicative of a small emission region, which also explains the relative lack of profile evolution with frequency.
However, since the number of MSPs with flux density measurements over a wide frequency range is small, the apparent differences to non-recycled pulsar spectra are not yet conclusive.

Low-frequency observations of MSPs ($\nu\lesssim\qty{300}{\MHz}$) present numerous challenges.
For one, the diffuse Galactic continuum emission scales with $\nu^{-2.55}$ \citep{Guzman2011}, which is steeper than most MSP spectra, thus reducing the signal-to-noise ratio (\snr) at lower frequencies.
Cold-plasma dispersion ($\propto\mathrm{DM}\nu^{-2}$, where DM is the dispersion measure) and scattering \citep[$\propto\nu^{-4}$;][]{Bhat2004} in the IISM are also much more significant at low frequencies.
Dispersive smearing can be mitigated using finer frequency channels, however this requires a trade-off with time resolution \citep[][for example, were limited to a time resolution of \qty{128}{\us}]{Kuzmin2001}.
In order to observe fine structure in MSP profiles at low frequencies, it is therefore necessary to coherently dedisperse the data in order to remove the intrachannel dispersive smearing \citep{Hankins1971,HankinsRickett1975,Hankins2018}.
The pulse broadening imparted on pulsar signals by multipath scattering the IISM cannot be corrected so easily, and prevents the detection of pulsations when the characteristic pulse broadening time significantly exceeds the pulsar rotation period.
This limits phase-resolved detections of MSPs to lower DMs than non-recycled pulsars, owing to the scaling of pulse broadening with DM \citep[e.g. see Equation~7 and Figure~4 of][]{Bhat2004}.
Since dispersion and scattering are greater over larger distances, they limit the maximum distance of pulsar detections at low frequencies.

The largest census of MSPs below \qty{300}{\MHz} to date was carried out with the Low-Frequency Array (LOFAR), where 48 MSPs were detected between \numrange{110}{188}\,\unit{MHz} at declinations north of $\delta\sim\qty{-30}{\degree}$ \citep{Kondratiev2016}.
The LOFAR observations enabled analysis of pulse profile evolution over a wide frequency range for a large sample of MSPs, providing further evidence that MSP profile widths evolve slowly with frequency whilst the individual component amplitudes can change dramatically.
A southern-sky MSP census below \qty{300}{\MHz} is necessary to provide a more complete catalogue of MSP properties at low frequencies, which will be an important resource for planning future pulsar surveys with the SKA-Low telescope.

The Murchison Widefield Array \citep[MWA;][]{Tingay2013} is a low-frequency (\numrange{80}{300}\,\unit{\MHz}) aperture array deployed at Inyarrimanha Ilgari Bundara, the CSIRO Murchison Radio-astronomy Observatory (MRO) in Western Australia, and is a key pathfinder for the SKA-Low.
The MWA is equipped with a high-time-resolution data recorder, the Voltage Capture System \citep[VCS;][]{Tremblay2015}, which has enabled detailed studies of several MSPs, including PSRs J0437$-$4715, J2145$-$0750, and J2241$-$5236 \citep{Bhat2014,Bhat2016,Bhat2018,Kaur2019,Kaur2022}.
The Southern-sky MWA Rapid Two-metre (SMART) survey exploits the flexibility of the VCS data to observe the entire sky south of declination $\delta\sim+\qty{30}{\degree}$ in less than 100 telescope-hours while still reaching a dwell time of $\sim$\qty{80}{\min}.
The survey is described in \citet{Bhat2023I} and the initial census and pulsar discoveries are described in \citet{Bhat2023II}.
In total, 15 MSP detections were reported in \citet{Bhat2023I}, most of which were made in search-mode data (detected Stokes timeseries with \qty{100}{\us} time resolution) as a byproduct of the first-pass SMART survey processing.
As a result, these data are incoherently dedispersed with integration times of \qty{10}{\min}.

Building on this previous work with greater time resolution and sensitivity, we undertook a full census of MSPs in the SMART data, including the final SMART observations covering the Galactic plane.
This paper describes our analysis and results, and is organised as follows.
In Section~\ref{sec:observations}, we describe the observations, target selection, and data reduction workflow.
In Section~\ref{sec:results}, we present and discuss the results of the census, including mean flux densities, profile-averaged and phase-resolved rotation measures, and a comparison of the integrated pulse profiles with higher frequency profiles in the published literature.
In Section~\ref{sec:future}, we discuss the limitations of the current work and the future outlook for high-time-resolution pulsar science with the MWA.
Finally, in Section~\ref{sec:conclusions}, we summarise the results of the census and present our conclusions.
    \section{Observations and data reduction}\label{sec:observations}

\subsection{Data acquisition and beamforming}
For this work, we exclusively used observations from the SMART pulsar survey.
We refer the reader to \citet{Bhat2023I} for a complete description of the signal chain and data acquisition, and provide a summary of the relevant details here.
The MWA comprises 256 tiles, with each tile being a $4\times 4$ array of dual-polarisation dipole antennas.
The SMART observations were collected between September 2018 and November 2023 in the Phase II compact configuration \citep[$\sim$128 tiles;][]{Wayth2018}.
All SMART observations were made over a contiguous \qty{30.72}{\MHz} wide frequency band centred at \qty{154.24}{\MHz}.
The legacy VCS \citep[in use before September 2021;][]{Tremblay2015} recorded $4+4$-bit complex voltage samples for each tile in $3072\times 10$-kHz `fine' channels with a total data rate of \qty{28}{\TB\per\h}.
The successor to the legacy VCS is the MWAX VCS \citep{Morrison2023}, which records $8+8$-bit complex voltage samples in $24\times 1.28$-MHz `coarse' channels with a total data rate of \qty{56}{\TB\per\h}.
VCS-recorded data are transferred to the Pawsey Supercomputing Centre for long-term storage and offline processing.

Each SMART observation was calibrated using single dedicated observation of a nearby calibrator source before or after the VCS observation.
We used \textsc{birli}\footnote{\url{https://github.com/MWATelescope/Birli}} to preprocess the correlator visibilities, which included downsampling to a frequency/time resolution of \qty{40}{\kHz}/\qty{2}{\s}.
We then used \textsc{hyperdrive}\footnote{\url{https://github.com/MWATelescope/mwa_hyperdrive}} to perform direction-independent `sky model' calibration, which yielded instrumental gains for each tile and frequency channel.
We typically flagged \qty{80}{\kHz} at the top and bottom edges of each receiver channel, leaving an effective bandwidth of \qty{26.88}{\MHz}.

The raw tile voltages were coherently combined using \textsc{vcsbeam}\footnote{\url{https://github.com/CIRA-Pulsars-and-Transients-Group/vcsbeam}}, a GPU-accelerated offline beamformer (\cite{Ord2019}; Bhat et al. in prep.).
As the beamformer is designed to operate on fine-channelised voltages, the coarse-channelised MWAX VCS data were passed through an offline polyphase filterbank to replicate the legacy VCS fine-channelisation.
The sensitivity of the tied-array beam is a strong function of look direction, with the greatest sensitivity being within the primary beam (field of view $\sim610\,\mathrm{deg}^2$ near zenith at 154\,MHz).
For standard pulsar searching and observations of non-recycled pulsars, 10-kHz/100-$\mu$s resolution and incoherent dedispersion is often sufficient; in these cases, the beamformer can output the detected Stokes intensities in the \textsc{psrfits} file format \citep{Hotan2004}.
For applications that require higher time resolutions or access to the complex voltages, the beamformed data are passed through a polyphase synthesis filter which reconstructs the 1.28-MHz resolution coarse channels, with a maximum time resolution of \qty{781}{\ns} \citep{McSweeney2020}.
The complex voltages are output in the VLBI Data Interchange Format (VDIF)\footnote{\url{https://vlbi.org/vlbi-standards/vdif/}}.

\subsection{Selection and location of targets}
Our aim was to observe a large sample of MSPs in order to obtain as many detections as possible in the SMART data.
As a starting point, our target list included the 88 MSPs regularly observed with the MeerTime Pulsar Timing Array \citep[MPTA;][]{Spiewak2022}.
We also included several pulsars from the LOFAR census of MSPs \citep{Kondratiev2016}, and all MSPs previously detected with the MWA \citep{Bhat2023II}.

\begin{figure}[t]
    \centering
    \includegraphics[width=\linewidth]{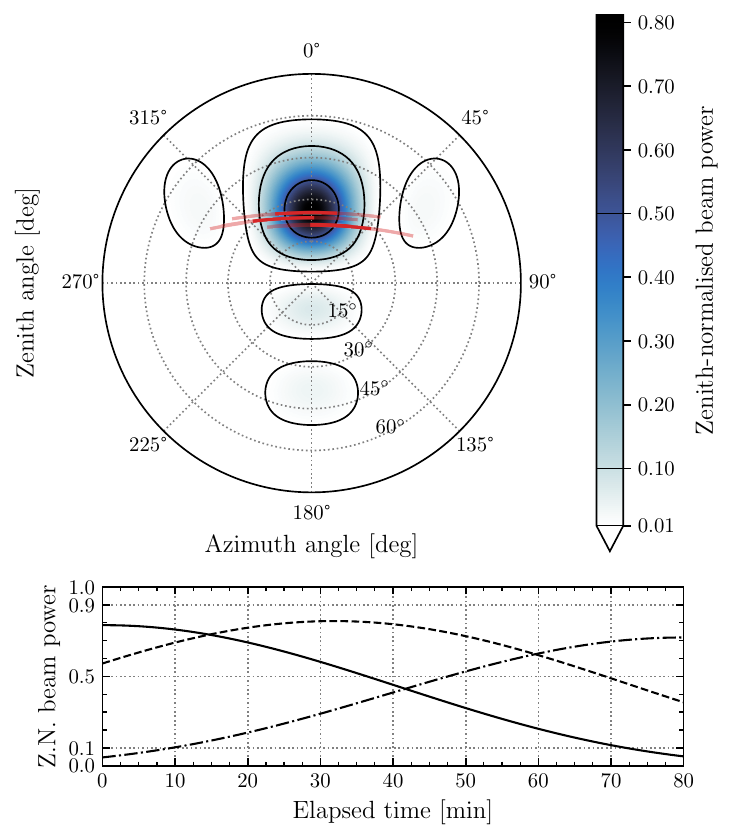}
    \caption{Paths of three pulsars through the primary beam during an 80-min SMART observation. Top: Polar plot displaying the zenith-normalised primary beam power at 154.24\,MHz in horizontal sky coordinates, with contours at the 1\%, 10\%, and 50\% power levels. The pulsars trace a path anticlockwise through the beam. The paths during the observation are shown in opaque red, and the the paths in the 1\,h before and after the observation are shown in translucent red. Bottom: The zenith-normalised beam powers towards each pulsar as a function of time during the observation. Each pulsar peaks in beam power at a different time.}
    \label{fig:beam}
\end{figure}

To identify which targets are in each SMART observation, we used \textsc{hyperbeam}\footnote{\url{https://github.com/MWATelescope/mwa_hyperbeam}} to model the primary beam at \qty{154.24}{\MHz}.
Due to the strong direction dependence of the beam, the power towards a target can vary significantly over an observation.
As such, it is often not beneficial to process a full SMART observation, as the pulsar may only be at a significant power level for part of the time.
Figure~\ref{fig:beam} shows the paths of three MSPs through the primary beam during a SMART observation (top panel) and the zenith-normalised beam powers towards each source over time (bottom panel).
The large size of VCS data also motivates a more streamlined processing strategy, in which we only download and process the raw voltages for time intervals containing the targets at a significant power level.
We therefore computed the primary beam power towards each MSP as a function of time throughout each observation, and used this to determine the optimal time ranges to download.
We processed between \numrange{30}{80}\,min for each target.
Where possible, we also beamformed towards other MSPs in the downloaded data (i.e. MSPs which were not on the initial target list, but were convenient to observe).

In total, we attempted to detect 154 of the 496 known Galactic pulsars with $P<\qty{30}{\ms}$\footnote{\url{https://www.astro.umd.edu/~eferrara/pulsars/GalacticMSPs.txt} (last updated 2024-04-05)} and declinations $\delta<+\qty{30}{\degree}$.
If we exclude pulsars with estimated scattering timescales greater than their spin period \citep[calculated using the scattering-DM relation from][]{Bhat2004}, we attempted to detect 122 of the 245 possible targets (i.e. $\sim$50\%).
However, this is only a rough estimate of the scattering limit, as the model from \citet{Bhat2004} is based on measurements from the Galactic plane and has roughly an order of magnitude uncertainty.
We also observed PSR J2222$-$0137, a mildly-recycled pulsar that does not meet the period cutoff ($P=\qty{32.8}{\ms}$), but is nonetheless an important target due to being in a relativistic binary \citep[e.g.][]{Kramer2021MNRAS}.

\subsection{Pulsar data processing}\label{sec:processing}
The census was carried out in two stages.
In the first stage, we beamformed towards all targets and produced detected Stokes beams with the standard fine channelisation (10-kHz/100-$\mu$s resolution).
This involved processing $\sim$\qty{1.5}{\PB} of raw VCS data from 47 SMART observations.
We used \textsc{prepfold} from the \textsc{presto} software suite \citep{RansomPhD} to incoherently dedisperse and fold the beams, then perform a grid search around the nominal spin period, spin period derivative, and DM to optimise the detection significance.
Where possible, we used timing ephemerides from the MeerTime project.
Otherwise, we used ephemerides from the ATNF pulsar catalogue \citep{Manchester2005}.
The detections were identified visually from the \textsc{prepfold} diagnostic plots.
Once a full census had been performed, we re-beamformed on the detected targets to produce voltage beams with coarse channelisation (\qty{1.28}{\MHz}/\qty{781}{\ns}).
We used \textsc{dspsr} \citep{vanStraten2011} to form synthetic filterbanks with 160-kHz channels and concurrently perform coherent dedispersion to the nominal DM.
For high-\snr{} detections with sharp features (e.g. PSR J2241$-$5236), we also generated 320-kHz filterbanks to obtain a higher time resolution.
The filterbanks were then folded at the nominal spin period with 5-second subintegrations.
The number of phase bins was chosen as the largest power of two less than $P/\delta t$, where $P$ is the spin period and $\delta t$ is the sample time resolution, but not more than 1024.
If the \snr{} was too low, we then downsampled in phase by powers of two.
The folded observations were cleaned with \textsc{clfd} \citep{Morello2019}, which typically flagged $<1\%$ of the data.
We then used \textsc{pdmp} from \textsc{psrchive} \citep{vanStraten2012,Hotan2004} to perform a grid search in the period-DM parameter space, with a DM step size of \qty{0.001}{\per\cm\cubed\pc}.
For MSPs with a significant DM correction, we reprocessed the data with the updated DM.

The two-stage processing strategy was adopted due to a software bug that affected the VDIF output from the beamformer (which underwent a significant overhaul in 2021 as the MWA transitioned from the legacy to the MWAX backend).
As a result, the VDIF output mode was unavailable while the initial census was being carried out.
We discuss the effect of incoherent dedispersion on the detected sample in Section~\ref{sec:results-overview}.
    \section{Results and Discussion}\label{sec:results}

\begin{table*}
\centering
\caption{
Summary of the 40 detected MSPs.
For each MSP, we list the B/J name, the Galactic coordinates ($\textit{l}$, $\textit{b}$), the best spin period and DM from \textsc{pdmp}, the mean flux density at 154\,MHz (flux densities which may be underestimated due to scattering are indicated with an asterisk), and the binary system/companion type, where `WD' denotes a white dwarf companion of either Helium (He) or CO/ONeMg (CO) type, `NS' a neutron star companion, `BW' a black widow system (very low mass companion, $\textit{M}_\text{c}\ll \text{0.1}\,\textit{M}_\odot$), and `Pl' one or more planetary companions.
For the best detection of each MSP, we list the epoch, the MWA obs ID, the integration time ($\Delta \textit{t}_\text{obs}$), and the mean offset from the pointing centre.
We also indicate the detections from the WSRT \citep[115--175\,MHz;][]{Stappers2008}, the Large Phased Array (LPA) radio telescope in Pushchino \citep[narrowband at 102 and 110\,MHz;][]{Malofeev2000,Kuzmin2001}, the Long Wavelength Array \citep[LWA; 26--88\,MHz;][]{Dowell2013,Kumar2025}, and LOFAR \citep[110--188\,MHz;][]{Kondratiev2016,Bassa2017,Sanidas2019}.
The telescope names are abbreviated to the first two letters.
}
\label{tab:detections}
\begin{tabular}{lcccccccccccc}
\toprule
PSR & $\textit{l}$ & $\textit{b}$ & Period & DM & $\textit{S}_\text{mean}$ & Binary? & Epoch & Obs ID & $\Delta \textit{t}_\text{obs}$ & Offset & Detections \\
 & [\unit{\degree}] & [\unit{\degree}] & [ms] & [$\text{cm}^{-\text{3}}\,\text{pc}$] & [mJy] &  & [MJD] &  & [min] & [\unit{\degree}] &  \\
\midrule
J0030+0451 & 113.1 & $-$57.6 & 4.865 & 4.33210(34) & 93(28) & Isolated & 58774.6 & 1255444104 & 30 & 4.3 & LP,LW,LO \\
J0034$-$0534 & 111.5 & $-$68.1 & 1.877 & 13.76460(26) & 190(60) & He WD & 58774.6 & 1255444104 & 30 & 7.9 & WS,LP,LW,LO \\
J0125$-$2327 & 188.9 & $-$81.6 & 3.675 & 9.5932(5) & 20(6) & Binary & 58434.6 & 1226062160 & 30 & 5.6 & $-$- \\
J0407+1607 & 176.6 & $-$25.7 & 25.702 & 35.609(7) & 65(19) & He WD & 58764.8 & 1254594264 & 30 & 3.2 & LO \\
J0437$-$4715 & 253.4 & $-$42.0 & 5.757 & 2.64310(34) & 400(120) & He WD & 58757.8 & 1253991112 & 30 & 7.4 & $-$- \\
J0621+1002 & 200.6 & $-$2.0 & 28.854 & 36.568(26) & 44(18)$^*$ & CO WD & 58823.7 & 1259685792 & 30 & 8.9 & LP,LO \\
J0711$-$6830 & 279.5 & $-$23.3 & 5.491 & 18.401(6) & 13(4)$^*$ & Isolated & 58907.6 & 1266932744 & 30 & 6.6 & $-$- \\
J0737$-$3039A & 245.2 & $-$4.5 & 22.699 & 48.9097(15) & 130(50)$^*$ & NS & 58841.7 & 1261241272 & 30 & 4.4 & LO \\
J0952$-$0607 & 243.7 & 35.4 & 1.414 & 22.4115(6) & 17(5) & BW & 58883.7 & 1264867416 & 30 & 8.1 & LO \\
J1022+1001 & 231.8 & 51.1 & 16.453 & 10.2524(8) & 46(14) & CO WD & 58909.7 & 1267111608 & 30 & 9.0 & WS,LP,LW,LO \\
J1024$-$0719 & 251.7 & 40.5 & 5.162 & 6.4814(21) & 16(5) & Isolated & 58911.7 & 1267283936 & 30 & 5.9 & LP,LO \\
J1038+0032 & 247.2 & 48.5 & 28.852 & 26.329(30) & 19(6)$^*$ & Isolated & 58920.7 & 1268063336 & 30 & 8.5 & LO \\
J1231$-$1411 & 295.5 & 48.4 & 3.684 & 8.0913(12) & 21(6) & He WD & 59299.7 & 1300809400 & 30 & 2.4 & LO \\
B1257+12 & 311.3 & 75.4 & 6.219 & 10.1524(7) & 50(15) & Pl & 59301.7 & 1300981728 & 30 & 6.4 & WS,LP,LW,LO \\
J1400$-$1431 & 327.0 & 45.1 & 3.084 & 4.93410(32) & 47(14) & He WD & 59321.7 & 1302712864 & 30 & 6.2 & LW \\
J1455$-$3330 & 330.7 & 22.6 & 7.987 & 13.5697(4) & 130(40) & He WD & 59316.7 & 1302282040 & 30 & 7.0 & $-$- \\
J1536$-$4948 & 328.2 & 4.8 & 3.080 & 38.0043(17) & 250(100) & He WD & 60070.7 & 1367428632 & 60 & 9.6 & $-$- \\
J1543$-$5149 & 327.9 & 2.5 & 2.057 & 50.9868(9) & 220(90)$^*$ & He WD & 60070.7 & 1367428632 & 70 & 8.2 & $-$- \\
J1603$-$7202 & 316.6 & $-$14.5 & 14.842 & 38.045(14) & 52(16) & CO WD & 60069.7 & 1367342464 & 30 & 0.8 & $-$- \\
B1620$-$26 & 351.0 & 16.0 & 11.076 & 62.839(15) & 22(7) & He WD + Pl & 60113.6 & 1371131072 & 80 & 6.7 & LP \\
J1730$-$2304 & 3.1 & 6.0 & 8.123 & 9.6256(19) & 55(22) & Isolated & 60084.7 & 1368640168 & 30 & 4.7 & LP,LO \\
J1744$-$1134 & 14.8 & 9.2 & 4.075 & 3.1388(6) & 48(19) & Isolated & 60076.8 & 1367946928 & 30 & 4.7 & WS,LP,LO \\
J1757$-$5322 & 339.6 & $-$14.0 & 8.870 & 30.796(7) & 40(12) & CO WD & 60104.7 & 1370367808 & 30 & 8.5 & $-$- \\
J1804$-$2717 & 3.5 & $-$2.7 & 9.343 & 24.666(7) & 86(35) & He WD & 60084.8 & 1368640168 & 30 & 1.8 & $-$- \\
J1843$-$1113 & 22.1 & $-$3.4 & 1.846 & 59.9594(16) & 120(50)$^*$ & Isolated & 60094.8 & 1369505736 & 50 & 4.6 & $-$- \\
J1911$-$1114 & 25.1 & $-$9.6 & 3.626 & 30.9633(12) & 81(32) & He WD & 60094.8 & 1369505736 & 60 & 4.5 & WS,LP,LO \\
J1918$-$0642 & 30.0 & $-$9.1 & 7.646 & 26.598(8) & 39(15) & He WD & 60094.8 & 1369505736 & 40 & 7.0 & LO \\
B1937+21 & 57.5 & $-$0.3 & 1.558 & 71.020(4) & 71(28)$^*$ & Isolated & 60097.8 & 1369764224 & 30 & 3.6 & LP,LO \\
J1944+0907 & 47.2 & $-$7.4 & 5.185 & 24.348(9) & 45(18)$^*$ & Isolated & 60097.8 & 1369764224 & 30 & 9.8 & LO \\
B1957+20 & 59.2 & $-$4.7 & 1.607 & 29.1078(5) & 100(40)$^*$ & BW & 60097.8 & 1369764224 & 30 & 5.6 & WS,LO \\
J2039$-$3616 & 6.3 & $-$36.5 & 3.275 & 23.9657(11) & 16(5) & He WD & 60125.8 & 1372184672 & 30 & 5.8 & $-$- \\
J2051$-$0827 & 39.2 & $-$30.4 & 4.509 & 20.7222(4) & 61(18) & BW & 60118.8 & 1371581520 & 30 & 5.4 & LP,LO \\
J2129$-$5721 & 338.0 & $-$43.6 & 3.726 & 31.8482(9) & 11.2(3.4) & He WD & 58380.6 & 1221399680 & 80 & 5.2 & $-$- \\
J2145$-$0750 & 47.8 & $-$42.1 & 16.052 & 9.0054(14) & 500(150) & CO WD & 60118.8 & 1371581520 & 60 & 7.2 & WS,LP,LW,LO \\
J2150$-$0326 & 53.6 & $-$40.8 & 3.511 & 20.6744(10) & 22(7) & Binary & 58385.6 & 1221832280 & 50 & 8.3 & $-$- \\
J2222$-$0137 & 62.0 & $-$46.1 & 32.818 & 3.2714(9) & 27(8) & CO WD & 58385.6 & 1221832280 & 80 & 6.3 & LO \\
J2234+0944 & 76.3 & $-$40.4 & 3.627 & 17.8263(12) & 11.6(3.5) & BW & 58385.6 & 1221832280 & 80 & 10.0 & $-$- \\
J2241$-$5236 & 337.5 & $-$54.9 & 2.187 & 11.41120(19) & 82(24) & BW & 58380.6 & 1221399680 & 80 & 6.8 & $-$- \\
J2256$-$1024 & 59.2 & $-$58.3 & 2.295 & 13.77500(19) & 82(25) & BW & 58395.6 & 1222697776 & 30 & 4.4 & $-$- \\
J2317+1439 & 91.4 & $-$42.4 & 3.445 & 21.8985(6) & 21(6) & He WD & 58399.6 & 1223042480 & 30 & 4.4 & LP,LO \\
\bottomrule
\end{tabular}
\end{table*}

\begin{figure*}
    \centering
    \includegraphics[width=\linewidth]{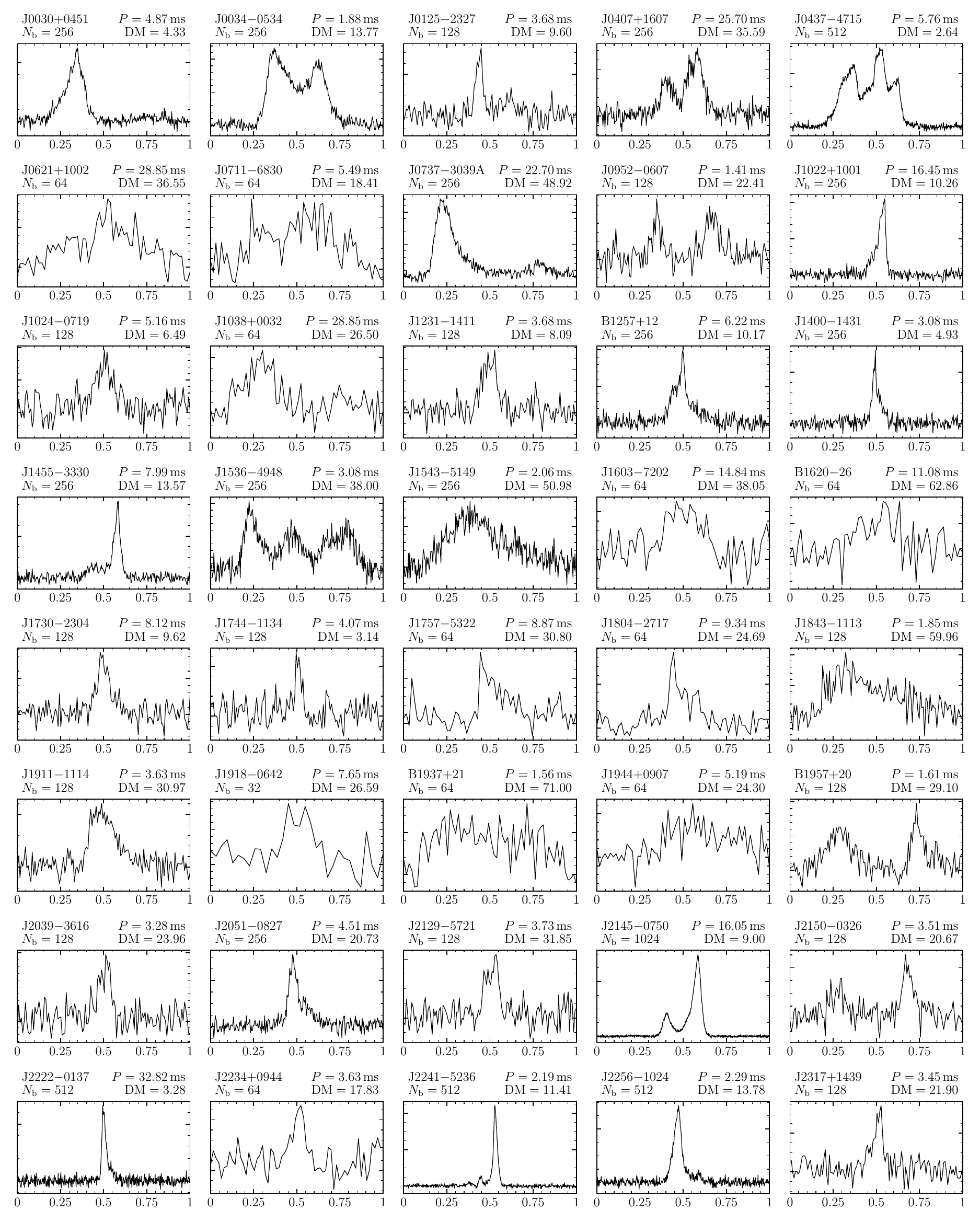}
    \caption{Integrated pulse profiles for the 40 detected MSPs in the frequency range 139--170\,MHz. For each pulsar, we list the number of phase bins ($\textit{N}_\text{b}$), the spin period ($\textit{P}$) in ms, and the DM in $\text{cm}^{-\text{3}}\,\text{pc}$.}
    \label{fig:grid}
\end{figure*}

\begin{figure*}[t]
    \centering
    \includegraphics[width=0.8\linewidth]{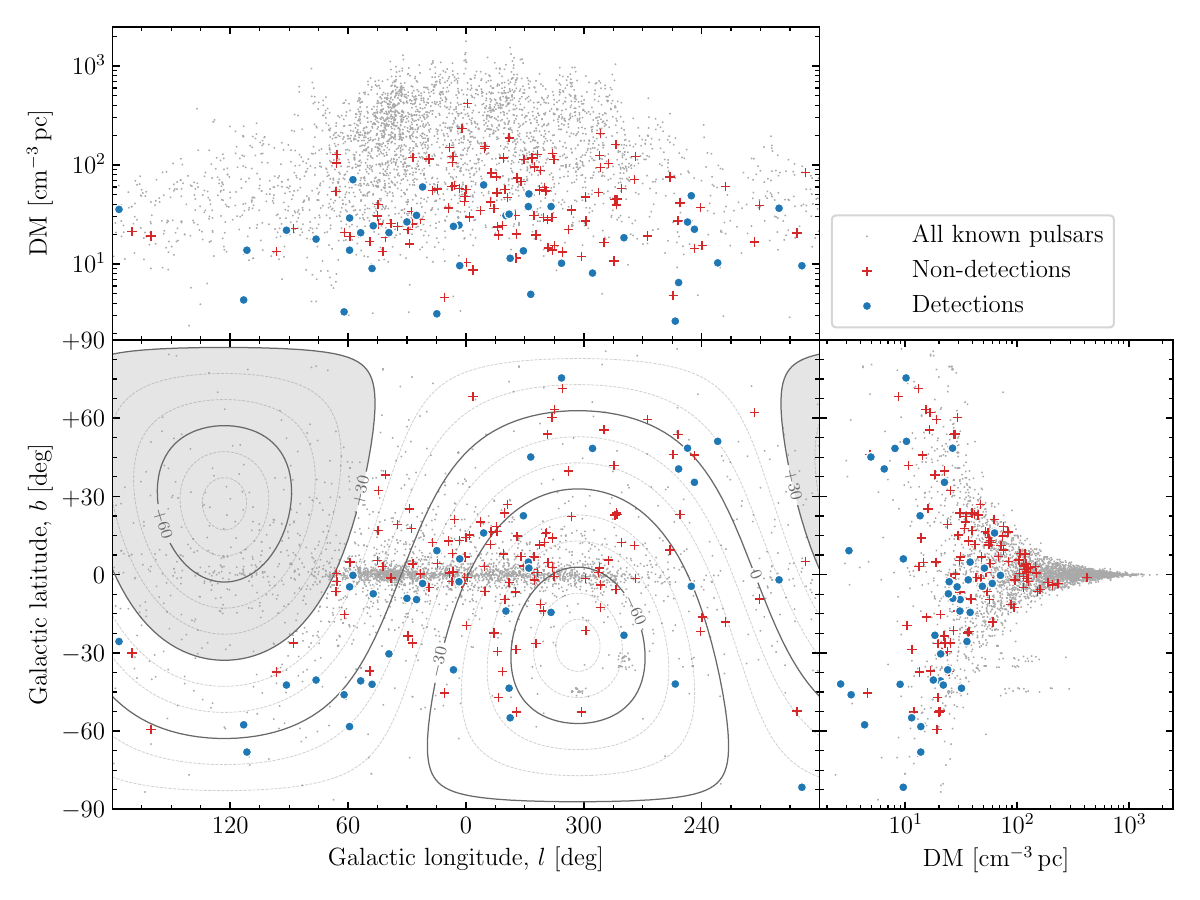}
    \caption{Galactic distribution of census targets. The detected targets are shown as blue filled circles and the non-detected targets are shown as red pluses. We also show all pulsars in the ATNF pulsar catalogue \citep[v2.6.1;][]{Manchester2005} as light grey dots. The contours indicate lines of equal declination and the shaded patches indicate the declinations out of reach of the MWA.}
    \label{fig:galdm}
\end{figure*}

\begin{figure}
    \centering
    \includegraphics[width=\linewidth]{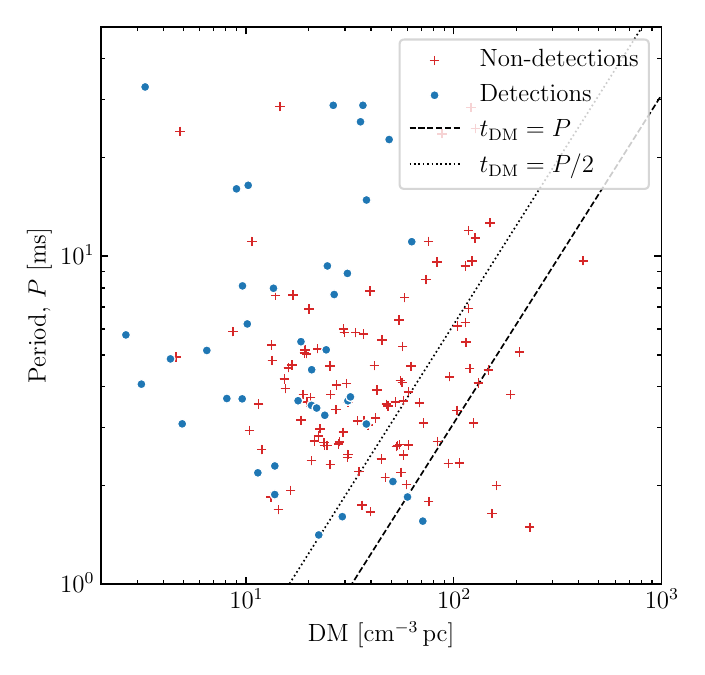}
    \caption{Distribution of census targets in the period-DM parameter space. The detected targets are shown as blue filled circles and the non-detected targets are shown as red pluses. As the initial census was incoherently dedispersed, we indicate where the intrachannel dispersive smearing ($\textit{t}_\text{DM}$) is equal to the period and half of the period with dashed and dotted lines, respectively.}
    \label{fig:pdm}
\end{figure}

\subsection{Overview}\label{sec:results-overview}
In total, we have detected 40 MSPs in SMART observations.
The detections are summarised in Table~\ref{tab:detections} and the integrated pulse profiles for each detection are showcased in Figure~\ref{fig:grid}.
As an indication of the expected sensitivity, we provide the mean offset of each target from the pointing centre (i.e. the phase centre of the primary beam) in Table~\ref{tab:detections}.
We also list whether each MSP has previously been detected (i.e. phase resolved) by other low-frequency telescopes.
We have detected 16 MSPs that have not previously been detected by other low-frequency telescopes.
Of these, 5 have had detections reported in previous MWA publications \citep[PSRs J0437$-$4715, J1455$-$3330, J1536$-$4948, J2241$-$5236, and J2256$-$1024; e.g.][]{Bhat2023II}.
The remaining 11 are the first published detections below \qty{300}{\MHz}.
The detected sample includes four MSPs in relativistic binaries \citep[PSRs J0737$-$3039A, J1603$-$7202, J1757$-$5322, and J2222$-$0137; see][]{Kramer2021MNRAS}, six black widow MSPs \citep[PSRs J0952$-$0607, B1957+20, J2051$-$0827, J2234+0944, J2241$-$5236, and J2256$-$1024; see][]{Koljonen2025}, and two MSPs with planets (PSRs B1257+12 and B1620$-$26).


Figure~\ref{fig:galdm} shows the Galactic distribution of the detected MSPs.
Given the relatively shallow reach of the census in DM, it is not surprising that the distribution of MSPs shows no bias towards the Galactic plane.
This is clearly demonstrated in the right panel of Figure~\ref{fig:galdm}, which shows the DM distribution as a function of Galactic latitude.

Figure~\ref{fig:pdm} shows the census sample as a function of period and DM, and indicates the line where the intrachannel dispersive smearing ($t_\mathrm{DM}$) exceeds the pulse period ($P$).
Pulsars to the right of the $t_\mathrm{DM}=P$ line on the diagram are expected to be completely smeared in the SMART frequency band, given the \qty{10}{\kHz} channelisation of the \textsc{psrfits} data.
These targets will need to be coherently dedispersed to confirm their detectability.
We note that PSR B1937+21 lies to the right of the $t_\mathrm{DM}=P$ line, and therefore was not detected in the initial census.
After performing coherent dedispersion, we were able to confirm a detection.
Due to the resources required to reprocess a large number of targets in VCS data (and particularly considering that the targets are spread throughout multiple SMART observations), we defer a full reprocessing of these targets to a future publication.

\subsection{Flux densities}
\subsubsection{Calibration}\label{sec:fluxcal}
The method used to calibrate MWA-VCS flux densities is described in \citet{Meyers2017}.
To summarise, the system temperature ($T_\mathrm{sys}$) is a combination of the antenna temperature ($T_\mathrm{ant}$), the receiver temperature ($T_\mathrm{rec}$), and the ambient temperature ($T_\mathrm{amb}$):
\begin{equation}
    T_\mathrm{sys} = \eta T_\mathrm{ant} + (1 - \eta) T_\mathrm{amb} + T_\mathrm{rec},
\end{equation}
where $\eta\approx 1$ is the radiation efficiency of the array.
To compute the antenna temperature, we perform an integral of the sky temperature weighted by the tied-array beam power over the sky.
The sky temperature was determined by scaling the 408-MHz Haslam map \citep[2014 reprocessed version;][]{Haslam1982,Remazeilles2015} to our observing frequencies assuming a spectral index of $-2.55$ \citep{Guzman2011}.
The tied-array beam power can be expressed as the product of the primary beam power and the array factor\footnote{The array factor is the complex-valued far-field radiation pattern of the array. It encapsulates the phase information required to form the tied-array beam.} power pattern.
Given that the product being integrated depends on the primary beam power, we performed the integral for only the sky regions where the primary beam power exceeded 0.1\% of the power at zenith.

The tied-array gain is defined as $G=A_\mathrm{e}/2k_\mathrm{B}$, where $A_\mathrm{e}$ is the tied-array effective area and $k_\mathrm{B}$ is the Boltzmann constant.
The effective area depends on the square of the observing wavelength ($\lambda^2$) and the the beam solid angle, which is an integral of the array factor power pattern over the whole sky.

The System Equivalent Flux Density (SEFD) of the array was calculated as,
\begin{equation}
    \mathrm{SEFD} = f_\mathrm{c} \frac{T_\mathrm{sys}}{G},
\end{equation}
where $f_\mathrm{c}$ is the coherency factor, which quantifies the coherent beamforming efficiency.
The efficiency of the beam formation is affected by the accuracy of the antenna phasing, differences in ionospheric refraction between the calibration observation and the target observation, and many other factors \citep{Ord2019}.
Rather than measuring $f_\mathrm{c}$ for each observation, we assigned a value of 0.7, and factored this assumption into the uncertainties.
We computed the SEFD at four uniformly spaced time and frequency steps over the total observing time and frequency band (for a total of 16 samples per observation) and used the mean SEFD for calibrating the flux scale.

The data were then processed as follows.
Each pulsar observation was first averaged over all sub-integrations and frequency channels, and the total intensity (Stokes $I$) pulse profile was extracted.
The profiles were then normalised by the standard deviation of the off-pulse noise to convert them to \snr{} units.
We identified the off-pulse region by searching for the phase window in which the integrated flux density is minimal.
For pulsars without a clear off-pulse region due to scattering, we measured the noise from a profile dedispersed to a DM of \qty{500}{\per\cm\cubed\pc}, which is sufficient to fully disperse the signal over the pulse phase.
We then measured the mean \snr{} over the profile, $\langle \mathrm{S}/\mathrm{N} \rangle$, and calculated the mean flux density ($S_\mathrm{mean}$) using the radiometer equation:
\begin{equation}\label{eq:radiometer}
    S_\mathrm{mean} = \langle \mathrm{S}/\mathrm{N} \rangle \times \frac{\mathrm{SEFD}}{\sqrt{n_\mathrm{p}\Delta\nu  \Delta t_\mathrm{obs} N_\mathrm{b}^{-1}}},
\end{equation}
where $n_\mathrm{p}$ is the number of polarisations summed ($n_\mathrm{p}=2$ for our data), $\Delta\nu$ is the bandwidth, $\Delta t_\mathrm{obs}$ is the integration time, and $N_\mathrm{b}$ is the number of phase bins\footnote{This is equivalent to flux-calibrating the profile and then measuring the mean flux density.}.

We assign conservative uncertainties to account for the following factors that influence the measured flux densities:
\begin{description}
    \item[\normalfont\itshape Beam coherence:] Based on past measurements, we estimate that the coherency factor of will vary by no more than $\sim$20--30\% from the nominal value of 0.7.
    \item[\normalfont\itshape Galactic emission spectrum:] The spectral index of the Galactic continuum emission is typically estimated to be between $-2.5$ and $-2.6$ away from the Galactic plane, and flattens due to thermal absorption to between $-2.2$ and $-2.5$ on the Galactic plane \citep{Guzman2011,Eastwood2018}. Spectral indices derived between the 159-MHz Engineering Development Array 2\footnote{The EDA2 is an SKA-Low prototype station which uses the same antennas and is located at the same site as the MWA.} (EDA2) sky map and the reprocessed 408-MHz Haslam sky map are generally consistent with these results \citep{Kriele2022}. Given our assumed spectral index of $-2.55$, we expect an error of up to $\sim$5\% of our extrapolated $T_\mathrm{sky}$ at high Galactic latitudes ($|b|>\qty{10}{\degree}$), and up to $\sim$30\% on the Galactic plane ($|b|<\qty{10}{\degree}$). Given that the primary beam is much wider than the Galactic plane, the total error in the sky integral should always be lower than this.
    \item[\normalfont\itshape Beam jitter:] Ionospheric refraction can cause beam `jitter' (i.e. dephasing) on timescales of $\sim$hours when the angular scale of the source position offsets (between the calibrator and the target) are comparable to the half width at half maximum (HWHM) of the tied-array beam. Observations from 2013--2016 (around Solar cycle maximum) show that ionospheric offsets are typically $\mathbin{\sim}0.1\text{--}0.2^\prime$ at the MRO \citep{Waszewski2022}.
    On rare occasions, extreme ionospheric activity (e.g. travelling ionospheric disturbances) can cause ionospheric shifts of up to $\mathbin{\sim}1^\prime$ \citep{Loi2015,Jordan2017}.
    The SMART observations were collected in 2018--2023 (at a comparable or lower Solar activity level to 2013--2016\footnote{\url{https://www.swpc.noaa.gov/products/solar-cycle-progression}}) in the compact configuration of the MWA, with an effective tied-array beam HWHM of $\mathbin{\sim}\numrange{10}{15}^\prime$ at \qty{154}{\MHz}.
    Therefore, even in extreme circumstances, we expect the beam power attenuation to be negligible ($<1\%$).
    Furthermore, most ionospheric activity evolves on time scales longer than the duration of the SMART observations, so the calibration solutions are likely an accurate representation of the ionospheric state, even during active periods.
\end{description}
To account for the cumulative error of the above factors, we assign a flux density uncertainty of 30\% to pulsars at $|b|>\qty{10}{\degree}$, and 40\% to pulsars at $|b|<\qty{10}{\degree}$.
Several other factors remain unaccounted because they are difficult or impossible to quantify.
These are:
\begin{description}
    \item[\normalfont\itshape Beam simulation:] The array factor calculation assumes an array of identical elements, which is not necessarily true for the MWA due to some tiles having one or more failing dipoles. Whilst we do not account for this, it is not likely to be a dominant error.
    \item[\normalfont\itshape Interstellar scintillation:] Low-DM pulsars ($\mathrm{DM}\lesssim\qty{10}{\per\cm\cubed\pc}$) are susceptible to diffractive interstellar scintillation (DISS) with characteristic timescales of $\mathbin{\sim}10\,\mathrm{min}$ and bandwidths of $\mathbin{\sim}\qty{1}{\MHz}$ \citep[e.g. PSR J0437$-$4715;][]{Bhat2016}. DISS has been observed to cause flux density variations of $\sim$5--6 for J0437$-$4715. Additionally, refractive interstellar scintillation (RISS) can cause variations on timescales of weeks to months. Quantifying DISS and RISS for each pulsar requires long-term monitoring on timescales much larger than the characteristic RISS timescale, which is beyond the scope of this census.
    \item[\normalfont\itshape Interstellar scattering:] For significantly scattered pulsars (where the scattering timescale is comparable to or exceeds the pulse period), our measurements may underestimate the pulsar flux density owing to the pulsar signal contaminating our estimate of the off-pulse noise. We note which flux densities are likely to be underestimated in Table~\ref{tab:detections}.
\end{description}

\begin{figure*}
    \centering
    \includegraphics[width=\linewidth]{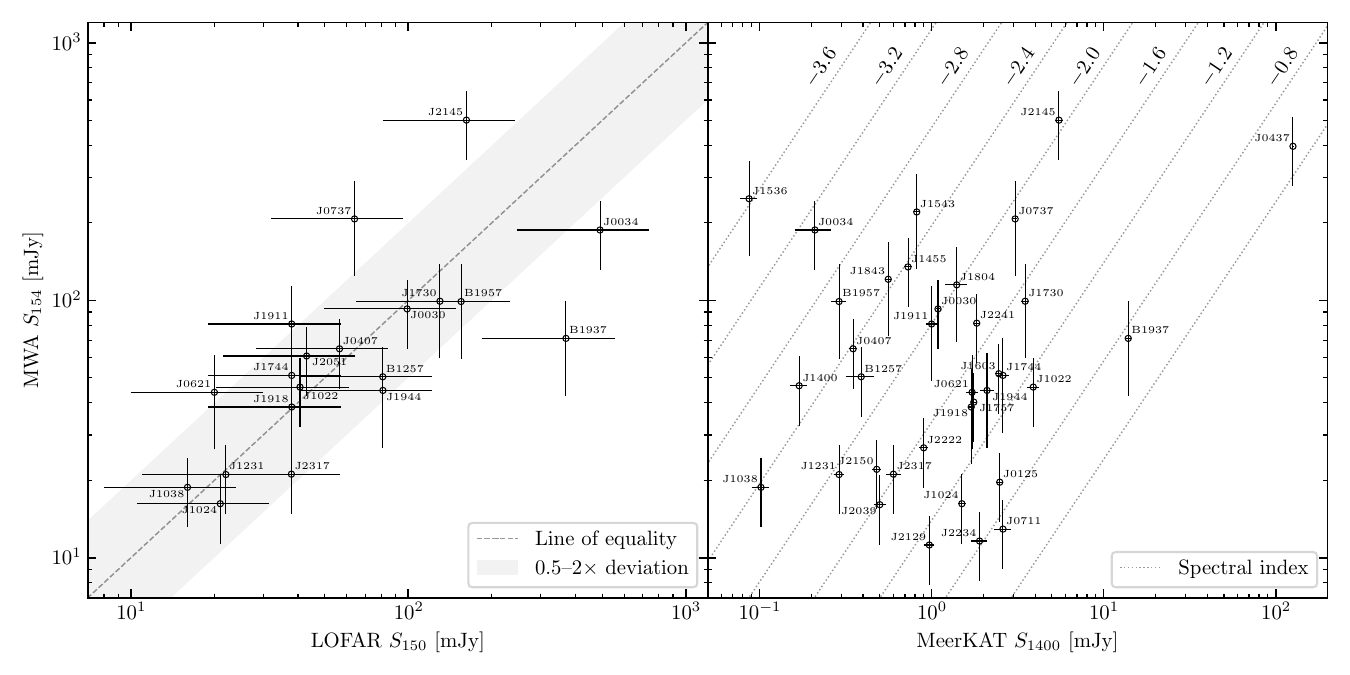}
    \caption{Comparison of MWA flux densities with LOFAR \citep[left panel; from][]{Kondratiev2016} and MeerKAT \citep[right panel; from][]{Spiewak2022}. Pulsars are labelled by right ascension. For the LOFAR comparison, we indicate the line of equal fluxes and the factor-of-two deviation band. For the MeerKAT comparison, we indicate the implied spectral index of a power-law model between the MWA and MeerKAT centre frequencies.}
    \label{fig:flux-compare}
\end{figure*}

\subsubsection{Comparison with LOFAR and MeerKAT}
We present mean flux densities for all of the detected MSPs in Table~\ref{tab:detections}.
Since our flux density calibration is based on a priori assumptions about the telescope and sky temperature, it is important to validate our measurements against independent measurements from other telescopes.
In the left panel of Figure~\ref{fig:flux-compare}, we compare flux densities for the MSP detections in common with LOFAR.
We see good agreement, with nearly all of the MWA flux densities comfortably falling within a factor of two of the LOFAR flux densities.
There are three notable outliers: PSRs J0737$-$3039A, B1937+21, and J2145$-$0750.
Both J0737$-$3039A and B1937+21 experience significant broadening due to interstellar scattering below \qty{200}{\MHz}.
Therefore, the measurements will have mismatched systematic errors due to the differing receiver bandwidths and contamination of the off-pulse noise estimate with the pulsar signal.
Radio continuum images of B1937+21 indicate a flux density of $\sim$1--2\,Jy in the MWA frequency band \citep{Erickson1985,Kuniyoshi2015}, however these may be overestimated due to source confusion.
Our flux density for J2145$-$0750 is consistent with measurements from the Engineering Development Array (EDA) in the same frequency band, which are averaged over 9 observations spanning 100 days \citep{Bhat2018}.
The discrepancy with LOFAR is most likely due to refractive interstellar scintillation.

In the right panel of Figure~\ref{fig:flux-compare}, we plot the MWA measurements against the MPTA census flux densities.
We indicate on the plot the spectral index required to account for the difference in flux density between the centre frequencies of the two telescopes, assuming a simple power-law model.
Due to the unaccounted uncertainties in our measurements (especially the intrinsic variability due to scintillation), the implied spectral indices are only an indicative value.
We also caution that although MSP spectra are typically described adequately by a simple power-law, spectral breaks (with flattening at lower frequencies) have been well constrained in some cases \citep[e.g. PSRs J0437$-$4715, B1937+21, and J2145$-$0750;][]{Kuniyoshi2015,Lee2022}.
In these cases, the spectral index implied by our comparison may be too steep at lower frequencies and too shallow at higher frequencies.
Regardless, the distribution of spectral indices we observe is consistent with previous MSP population studies, which find that the mean spectral index is around $-1.6$ to $-1.9$ \citep[e.g.][]{Toscano1998,Kramer1999,NickThesis,Karastergiou2024}.

The shallowest spectral index in the compared sample is for PSR J0437$-$4715, being $-0.52^{+0.17}_{-0.13}$.
This is shallower than the value of $-0.83$ expected for this frequency range from spectral modelling of published flux densities \citep{Lee2022}.
The discrepancy can be accounted for by the fact that J0437$-$4715 is known to exhibit strong diffractive scintillation at MWA frequencies \citep{Bhat2014,Bhat2018}, and the observation appears to have been taken when the pulsar was scintillating down.
The steepest spectral index is for PSR J1536$-$4948, being $-3.61^{+0.28}_{-0.21}$.
This is consistent with the steep in-band spectral index of $-3.2\pm 0.6$ from MeerKAT \citep{Spiewak2022}.
A more detailed spectral analysis using a larger set of flux densities compiled from the literature is deferred to a future publication.


\subsection{Rotation measures}
\subsubsection{RM synthesis}
We can describe the complex linear polarisation in terms of the Stokes parameters $I$, $Q$, and $U$ as
\begin{equation}
    \boldsymbol{P} = \boldsymbol{p}I = Q + \mathrm{i}U,
\end{equation}
where $\boldsymbol{p} \equiv \boldsymbol{P}/I$.
Written as a complex exponential, this is
\begin{equation}
    \boldsymbol{P} = \lVert \boldsymbol{p} \rVert I\mathrm{e}^{2\mathrm{i}\psi},
\end{equation}
where $\lVert \boldsymbol{p} \rVert$ is the degree of linear polarisation and $\psi$ is the position angle of the polarisation vector.
The linearly polarised intensity, $L$, is obtained by adding $Q$ and $U$ in quadrature:
\begin{equation}
    L \equiv \lVert \boldsymbol{p} \rVert I = \sqrt{Q^2 +U^2}.
\end{equation}
Polarised signals propagating through magnetised plasma experience a rotation of the polarisation vector due to the Faraday effect.
At an observing wavelength $\lambda$, the measured polarisation angle is
\begin{equation}\label{eq:lambda2}
    \psi = \psi_0 + \mathrm{RM}\lambda^2,
\end{equation}
where $\psi_0$ is the intrinsic polarisation angle and the rotation measure (RM) is the rate of change of the polarisation angle as a function of $\lambda^2$, usually in units of \unit{\radian\per\m\squared}.
Physically, the RM is a measure of the integrated magnetic field projected along the line of sight, weighted by the number density of free electrons.
Equation~\eqref{eq:lambda2} is valid assuming that the linear polarisation originates from a single source with no internal Faraday rotation.
It is typically assumed that Faraday rotation within the pulsar magnetosphere is negligible, which is theoretically supported under specific magnetospheric conditions \citep[][]{Wang2011}.
Therefore, unambiguous evidence for magnetospheric Faraday rotation (e.g. RM variations as a function of pulse phase or deviations from linearity in $\lambda^2$) can provide valuable constraints on magnetospheric models.
As such, it is important to consider these effects when measuring the RM from pulsar observations.

The simplest way to measure the RM is to fit Equation~\eqref{eq:lambda2} to the position angle as a function of $\lambda^2$.
However, we instead chose to use the technique of RM synthesis developed by \citet{Burn1966,Brentjens2005}, which has been widely adopted due to its superior sensitivity when analysing low-\snr{} data and its ability to separate instrumental polarisation from astrophysical signals.
Following \citet{Burn1966}, we consider a generalisation of the RM, known as the Faraday depth ($\phi$), which is defined at every point along the line of sight to the source.
Assuming $\psi_0$ is the same at all Faraday depths, we can define the Faraday dispersion function (FDF), which represents the intrinsic linear polarisation at each Faraday depth integrated over all $\lambda^2$:
\begin{equation}
    \boldsymbol{F}(\phi) = \int_{-\infty}^{+\infty} \boldsymbol{P}(\lambda^2)\, \mathrm{e}^{-2\mathrm{i}\phi\lambda^2}\,\mathrm{d}\lambda^2.
\end{equation}
If the polarised signal is negligibly Faraday rotated by the pulsar magnetosphere, then it will appear as a compact source in the FDF, with the amplitude peaking at the RM defined by Equation~\eqref{eq:lambda2}.
Hence, any excess signal in the FDF other than the RM peak implies some deviation from the $\lambda^2$ relation (which may be astrophysical or instrumental in nature).
In particular, weakly-chromatic or achromatic instrumental effects can appear as a `zero peak' in the FDF (i.e. a peak at $\phi\sim\qty{0}{\radian\per\m\squared}$).
If the RM peak can be resolved from the zero peak, then the RM can be measured without contamination from the instrumental signal.

As we can only observe $\boldsymbol{P}(\lambda^2)$ at discrete and positive values of $\lambda^2$, the practical implementation is a discrete Fourier transform,
\begin{equation}\label{eq:fdf}
    \tilde{\boldsymbol{F}}(\phi) \approx \frac{1}{N_\mathrm{c}} \sum_{c=1}^{N_\mathrm{c}} \boldsymbol{P} (\lambda^2_c)\, \mathrm{e}^{-2\mathrm{i}\phi (\lambda_c^2-\lambda_0^2)},
\end{equation}
where $c$ is the frequency channel index, $N_\mathrm{c}$ is the number of channels, $\lambda_c$ is the centre wavelength of channel $c$, and $\lambda_0$ is a reference wavelength to which all polarisation vectors are de-Faraday rotated \citep{Brentjens2005}.
For this work, we are only concerned with the amplitude of the FDF, so the choice of $\lambda_0$ is arbitrary.

As shown by \citet{Brentjens2005}, the reconstructed FDF can be described as the convolution of the true FDF with the RM spread function (RMSF),
\begin{equation}
    \boldsymbol{R}(\phi) \approx \frac{1}{N_\mathrm{c}} \sum_{c=1}^{N_\mathrm{c}} \mathrm{e}^{-2\mathrm{i}\phi (\lambda_c^2-\lambda_0^2)}.
\end{equation}
The RMSF can be deconvolved from the reconstructed FDF using the \textsc{rm-clean} algorithm \citep{Heald2009}.
This reduces sidelobe confusion, which can help to distinguish the zero peak from the RM peak.
The \textsc{rm-clean} model components can also help to identify sources of polarisation in the FDF.
The effective resolution of the FDF is essentially limited by the width of the main peak of the RMSF, which we quantify with full width at half maximum (FWHM):
\begin{equation}
    \delta\phi\ \left[\unit{\radian\per\m\squared}\right] \approx \frac{3.8}{\Delta\lambda^2},
\end{equation}
where $\Delta\lambda^2$ is the total span of the data in $\lambda^2$ space in \unit{\m\squared} \citep{Schnitzeler2009}.
For the SMART frequency band, $\delta\phi\sim\qty{2.50}{\radian\per\m\squared}$.

\begin{table*}[t]
\centering
\caption{
RM measurements for 25 MSPs detected in the census.
For comparison, we list RM measurements from MeerKAT \citep[][labelled `sbm22']{Spiewak2022} and Murriyang/Parkes \citep[][labelled `dhm15']{Dai2015}, as well as the $\textit{z}$-score of the reference measurements relative to the MWA measurements, $\textit{z}_\text{RM}$ (see Equation~\ref{eq:rmresid}).
Observed RMs which may be contaminated by instrumental polarisation (i.e. $\left|\text{RM}_\text{obs}^\text{MWA}\right|<\delta\phi/\text{2}$), are indicated with an asterisk.
}
\label{tab:rm}
\begin{tabular}{lccccccc}
\toprule
PSR & $\text{RM}_\text{obs}^\text{MWA}$ & $\text{RM}_\text{iono}^\text{MWA}$ & $\text{RM}_\text{IISM}^\text{MWA}$ & $\text{RM}_\text{IISM}^\text{sbm22}$ & $\textit{z}_\text{RM}^\text{sbm22}$ & $\text{RM}_\text{IISM}^\text{dhm15}$ & $\textit{z}_\text{RM}^\text{dhm15}$ \\
 & [$\text{rad}\,\text{m}^{-\text{2}}$] & [$\text{rad}\,\text{m}^{-\text{2}}$] & [$\text{rad}\,\text{m}^{-\text{2}}$] & [$\text{rad}\,\text{m}^{-\text{2}}$] &  & [$\text{rad}\,\text{m}^{-\text{2}}$] &  \\
\midrule
J0030+0451 & 1.16(8)$^*$ & $-$0.87(16) & 2.03(18) & 1.32(13) & $-$3.19 & -- & -- \\
J0125$-$2327 & 4.98(14) & $-$0.81(9) & 5.79(17) & 4.78(7) & $-$5.57 & -- & -- \\
J0437$-$4715 & 0.746(9)$^*$ & $-$0.49(13) & 1.24(13) & 0.150(20) & $-$8.55 & 0.58(9) & $-$4.25 \\
J0711$-$6830 & 23.75(14) & $-$0.51(7) & 24.26(16) & 23.90(30) & $-$1.06 & 23.9(4) & $-$0.83 \\
J0737$-$3039A & 120.244(29) & $-$0.68(10) & 120.92(10) & 120.46(4) & $-$4.28 & -- & -- \\
J1022+1001 & 0.813(30)$^*$ & $-$1.09(14) & 1.90(15) & 1.83(5) & $-$0.47 & 2.40(10) & 2.79 \\
J1024$-$0719 & $-$2.9(4) & $-$0.74(12) & $-$2.2(4) & $-$2.89(14) & $-$1.46 & $-$2.40(20) & $-$0.40 \\
J1038+0032 & 16.29(18) & $-$0.86(13) & 17.15(22) & 20(5) & 0.57 & -- & -- \\
J1231$-$1411 & 11.18(25) & $-$0.88(6) & 12.06(25) & 11.60(20) & $-$1.44 & -- & -- \\
B1257+12 & 7.42(32) & $-$1.07(7) & 8.49(33) & 9.0(8) & 0.59 & -- & -- \\
J1400$-$1431 & 4.26(8) & $-$1.22(6) & 5.49(10) & 5.0(3.0) & $-$0.16 & -- & -- \\
J1455$-$3330 & 14.17(15) & $-$0.73(5) & 14.90(15) & 14.7(8) & $-$0.25 & -- & -- \\
J1730$-$2304 & $-$5.16(27) & $-$1.29(6) & $-$3.88(27) & $-$3.72(9) & 0.54 & $-$8.8(6) & $-$7.48 \\
J1744$-$1134 & 1.3(4) & $-$1.43(6) & 2.7(4) & 1.62(9) & $-$2.61 & 2.20(20) & $-$1.08 \\
J1911$-$1114 & $-$28.41(19) & $-$1.17(9) & $-$27.24(21) & $-$28.0(5) & $-$1.39 & -- & -- \\
B1937+21 & 7.26(14) & $-$1.80(18) & 9.06(23) & 7.80(20) & $-$4.16 & 8.30(10) & $-$3.06 \\
J2039$-$3616 & $-$15.29(21) & $-$1.06(5) & $-$14.23(21) & $-$14.5(4) & $-$0.60 & -- & -- \\
J2051$-$0827 & $-$32.88(8) & $-$1.14(13) & $-$31.74(15) & -- & -- & -- & -- \\
J2129$-$5721 & 19.93(14) & $-$0.52(7) & 20.45(16) & 22.70(20) & 8.81 & 22.30(30) & 5.46 \\
J2145$-$0750 & $-$2.170(10) & $-$1.09(17) & $-$1.08(17) & 0.3(5) & 2.62 & $-$0.80(10) & 1.44 \\
J2150$-$0326 & 6.76(16) & $-$0.95(13) & 7.71(20) & 7.7(5) & $-$0.02 & -- & -- \\
J2222$-$0137 & $-$0.30(15)$^*$ & $-$0.96(13) & 0.65(20) & 1.35(13) & 2.90 & -- & -- \\
J2241$-$5236 & 12.349(21) & $-$0.55(8) & 12.90(8) & 12.00(30) & $-$2.91 & 13.30(10) & 3.09 \\
J2256$-$1024 & 13.184(32) & $-$0.82(12) & 14.00(12) & -- & -- & -- & -- \\
J2317+1439 & $-$9.8(5) & $-$1.39(14) & $-$8.4(5) & $-$9.8(4) & $-$2.26 & -- & -- \\
\bottomrule
\end{tabular}
\end{table*}

\subsubsection{RM estimation and uncertainties}\label{sec:bootstrap}
As described by \citet{Brentjens2005}, the uncertainty in the RM can be estimated analytically.
However, this method relies on assumptions about the shape of the channel bandpass that often do not hold for real data \citep{Schnitzeler2015}.
It can also be difficult to estimate a reliable analytic uncertainty for noisy data.
We therefore adopt a bootstrapping approach, similar to the method used by \citet{Ilie2019}, as follows.
White noise was added to Stokes $Q$ and $U$ with a standard deviation determined by taking the median (over all off-pulse phase bins) of the phase-resolved standard deviations estimated over all frequency channels.
RM synthesis was then performed on the simulated data and the RM was measured by fitting a parabola to the peak of the FDF\footnote{As we are only interested in locating peak of the FDF to measure the RM, we did not apply \textsc{rm-clean} when bootstrapping.} and interpolating to find the maximum.
This was repeated for a large number of iterations ($\geq 10^4$) to create a Gaussian distribution of simulated RM measurements.
The RM measurement and statistical uncertainty were taken as the mean and standard deviation of the bootstrap distribution, respectively.
For some low-\snr{} detections, the bootstrap distribution was multimodal due to the noise in the FDF having a comparable amplitude to the RM peak.
In these cases, if the majority of samples converged on a single RM, then we discarded the outlier samples before measuring the distribution.

\subsubsection{Ionospheric RM correction}
The observed RM is the sum of contributions from the IISM and the terrestrial ionosphere:
\begin{equation}
    \mathrm{RM}_\mathrm{obs} = \mathrm{RM}_\mathrm{IISM} + \mathrm{RM}_\mathrm{iono}.
\end{equation}
To estimate the ionospheric RM, we used \textsc{spinifex} \citep{spinifex}.
The total electron content (TEC) was estimated using a single-layer ionospheric model with global ionosphere maps provided by the Jet Propulsion Laboratory \citep[JPL;][]{Martire2024}.
The JPL maps are created using measurements from the International GNSS Service ground station network\footnote{\url{https://network.igs.org/}}, including one station at the MRO.
\textsc{spinifex} performs spatial and temporal interpolation of the TEC, and calculates uncertainties using the root-mean-square ionosphere maps.
Since \textsc{spinifex} is essentially a re-write of \textsc{rmextract} \citep{rmextract}, this method is practically equivalent to the one validated by \citet{Lee2024} using pulsar observations at the MRO.
To estimate $\mathrm{RM}_\mathrm{iono}$, we calculated the mean of 10 time steps spanning the duration of each pulsar observation.
An indicative statistical uncertainty was estimated by bootstrapping $\mathrm{RM}_\mathrm{iono}$ and calculating the standard deviation; this was then doubled to bring the mean ionospheric uncertainty of our dataset to $\sim$\qty{0.1}{\radian\per\m\squared}, which is more reflective of the accuracy observed by \citet{Lee2024}.
We then subtracted $\mathrm{RM}_\mathrm{iono}$ from the observed RM to obtain the estimated contribution from the IISM.

\begin{figure}[t]
    \centering
    \includegraphics[width=\linewidth]{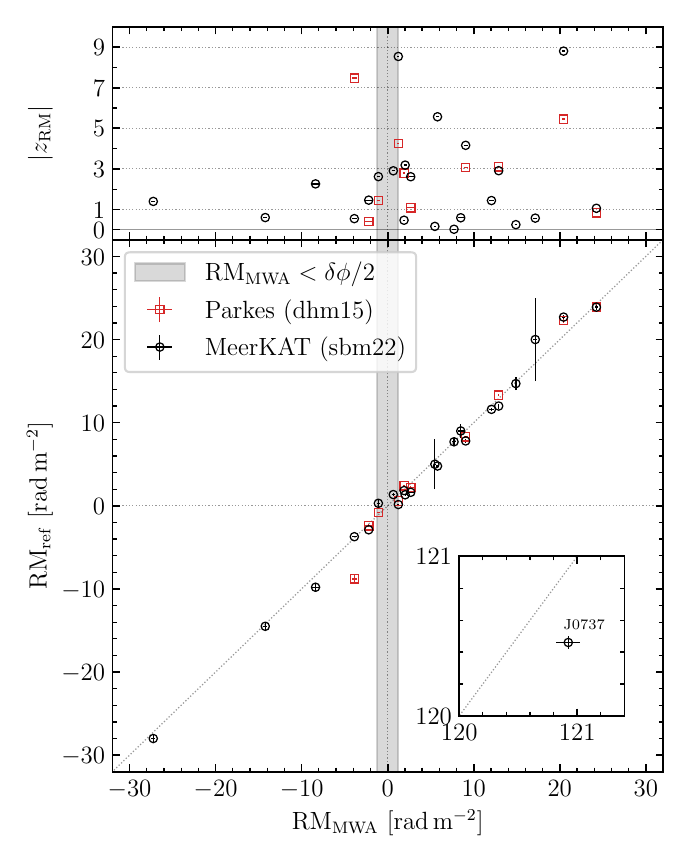}
    \caption{Comparison of the RM measurements from this work ($\text{RM}_\text{MWA}$) and published RM measurements from the literature ($\text{RM}_\text{ref}$). The red squares are Parkes measurements from \citet{Dai2015} and the black circles are MeerKAT measurements from \citet{Spiewak2022}. The shaded grey band shows the FWHM of the RMSF centred at $\text{0}\,\text{rad}\,\text{m}^{-\text{2}}$; MWA measurements within this range could potentially be contaminated with instrumental polarisation. For visual clarity, we show the measurement for PSR J0737$-$3039A in an inset. In the top panel, we show the absolute value of $\textit{z}_\text{RM}$ (see Equation~\ref{eq:rmresid}).}
    \label{fig:rm-compare}
\end{figure}

\begin{figure*}
    \centering
    \includegraphics[width=\linewidth]{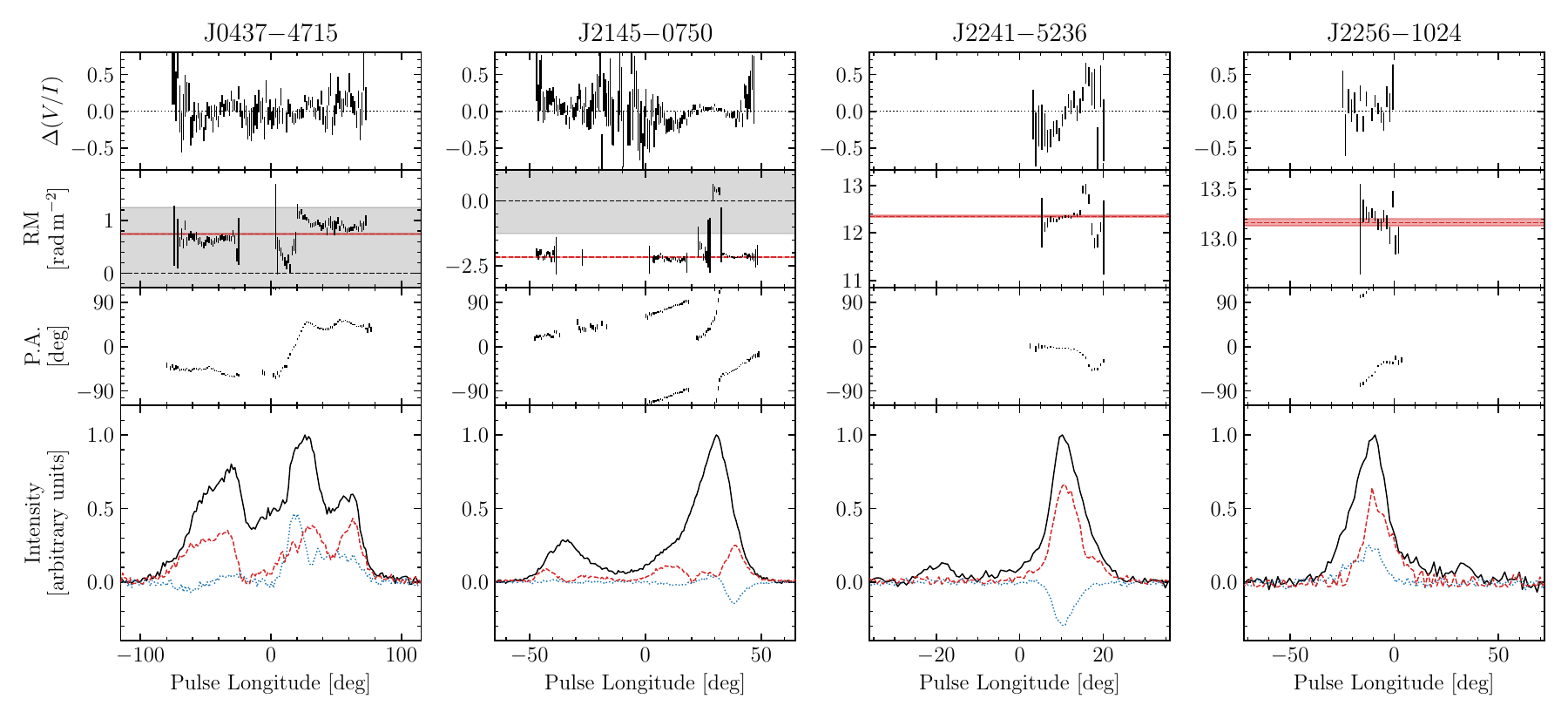}
    \caption{Polarimetric integrated pulse profiles for MSPs showing the most significant apparent phase-dependent RM variations. The bottom panels show the total intensity ($\textit{I}$; black solid line), linear polarisation ($\textit{L}$; red dashed line), and circular polarisation ($\textit{V}$; blue dotted line). The top panels show the change in circular polarisation over the observing band for bins with $\textit{I}>\text{10}\sigma_\textit{I}$, and the RM and position angle (P.A.) for bins with $\textit{L}>\text{5}\sigma_\textit{I}$. The red dashed lines and shaded RM ranges indicate the profile-averaged RMs and 1-$\sigma$ uncertainties as described in Section~\ref{sec:profile-rm}. The black dashed lines and shaded regions indicate the FWHM of the RMSF centred at $\text{0}\,\text{rad}\,\text{m}^{-\text{2}}$ (i.e. the zero peak).}
    \label{fig:pol-profs}
\end{figure*}

\subsubsection{Profile-averaged RM}\label{sec:profile-rm}
Following \citet{Ilie2019}, we determined a profile-averaged RM by summing together the amplitudes of all the on-pulse FDFs -- i.e.
\begin{equation}
    \tilde{F}_\mathrm{total}(\phi) = \sum_{\theta\in\Theta} \lVert \tilde{\boldsymbol{F}}_{\theta}(\phi) \rVert,
\end{equation}
where $\Theta$ is the set of all on-pulse phase bins -- and measuring the RM from the peak of $\tilde{F}_\mathrm{total}(\phi)$.
The on-pulse region was identified by searching for the phase window in which the integrated flux density was maximal.
The size of the phase window was chosen by eye to match the pulse width due to the large variation in pulse shapes and \snr{}.
The uncertainty was estimated by bootstrapping the profile-averaged RM following the method described in Section~\ref{sec:bootstrap}.

We obtained significant profile-averaged RM measurements for 25 MSPs, which are listed in Table~\ref{tab:rm} along with the estimated contributions from the ionosphere and IISM.
For comparison, we list the RM measurements from MeerKAT \citep{Spiewak2022} and Murriyang/Parkes \citep{Dai2015}, as well as the $z$-score of the reference measurements (from the literature) relative to the MWA measurements; i.e.
\begin{equation}\label{eq:rmresid}
    z_\mathrm{RM} = \frac{\mathrm{RM}_\mathrm{ref} - \mathrm{RM}_\mathrm{MWA}}{\sqrt{\sigma_\mathrm{ref}^2 + \sigma_\mathrm{MWA}^2}},
\end{equation}
where $\mathrm{RM}_\mathrm{ref} \pm \sigma_\mathrm{ref}$ are the reference measurements, and $\mathrm{RM}_\mathrm{MWA} \pm \sigma_\mathrm{MWA}$ are measurements from this work.
Figure~\ref{fig:rm-compare} shows a visual comparison of our measurements with the literature.
In general, our measurements are in good agreement with both MeerKAT and Parkes.
Of the 23 MSPs with both MWA and MeerKAT measurements, 17 are consistent to within $3\sigma$ and only three exceed $5\sigma$ (PSRs J0125$-$2327, J0437$-$4715, and J2129$-$5721).
Of the 10 MSPs with both MWA and Parkes measurements, 5 are consistent to within $3\sigma$ and two exceed $5\sigma$ (PSRs J1730$-$2304 and J2129$-$5721).
The noted discrepancies with MeerKAT and Parkes may be a result of different measurement methods.
For example, \citet{Dai2015} measured the RM for selected pulse phase regions in which the position angles were stable across the Parkes receiver bands.
This may also be the reason for the discrepancies between MeerKAT and Parkes; for J1730$-$2304, we observe a $7.48\sigma$ discrepancy with Parkes, but only a $0.54\sigma$ discrepancy with MeerKAT.
Overall, we find that the precision of our measurements is consistently comparable (and in some cases better) than the MeerKAT measurements (which are averages over $>6$ observations) and the Parkes measurements (which are fits over the full frequency range of the Parkes dual-band receiver).
Notably, PSRs J1038+0032 and J1400$-$1431 are depolarised at higher frequencies, and as a result the MWA measurements are more than an order of magnitude more precise than MeerKAT.

Two MSPs do not have RM measurements from MeerKAT or Parkes: PSRs J2051$-$0827 and J2256$-$1024.
For J2051$-$0827, the pulsar catalogue reports a value of \qty{-46 \pm 10}{\radian\per\m\squared} from \citet{Han2018}.
Our measurement of \qty{-31.74 \pm 0.15}{\radian\per\m\squared} is consistent within $1.5\sigma$ of the catalogued measurement and improves the precision by two orders of magnitude.
For J2256$-$1024, \citet{Crowter2020} report observed RM measurements (i.e. without ionospheric corrections) for three frequency bands: \qty{350}{\MHz}, \qty{820}{\MHz}, and \qty{1500}{\MHz}.
Our measurement of \qty{14.00 \pm 0.12}{\radian\per\m\squared} is consistent to within $1.2\sigma$ and $0.3\sigma$ of the \citeauthor{Crowter2020} measurements at \qty{820}{\MHz} and \qty{1500}{\MHz}, respectively.
The 350-MHz measurement is higher than ours by $8\sigma$; however, this could reasonably be attributed to the uncorrected ionospheric contribution.



\begin{figure*}
    \centering

    \begin{subfigure}[b]{0.49\textwidth}
        \centering
        \includegraphics[width=\textwidth]{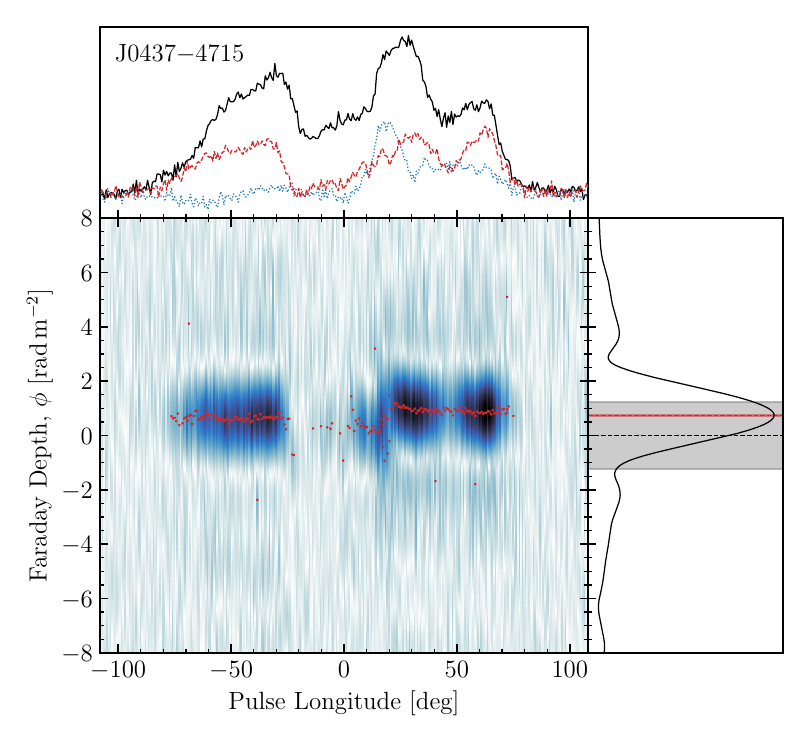}
    \end{subfigure}
    \hfill
    \begin{subfigure}[b]{0.49\textwidth}
        \centering
        \includegraphics[width=\textwidth]{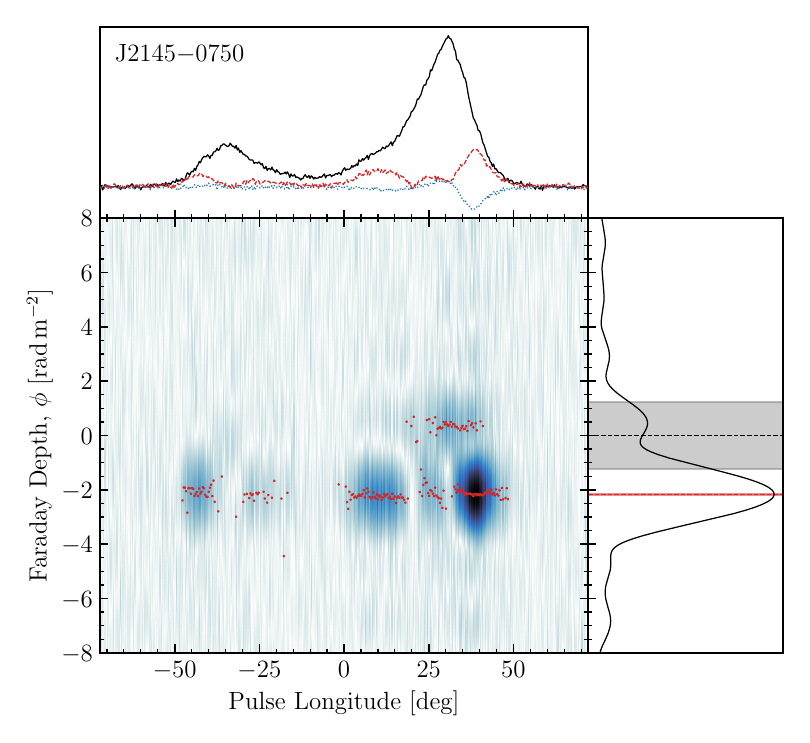}
    \end{subfigure}
    
    \caption{Phase-resolved and profile-averaged FDF amplitudes for PSRs J0437$-$4715 (left) and J2145$-$0750 (right). The top panels show the polarimetric integrated pulse profiles (as in Figure~\ref{fig:pol-profs}). The bottom left panels show the amplitude of the FDFs as a function of Faraday depth ($\phi$) and pulse longitude. The right panels show the average amplitudes of the on-pulse FDFs: the red dashed lines and shaded ranges indicate the RMs measured from the profile-averaged FDF amplitudes and their 1-$\sigma$ uncertainties; and the black dashed lines and shaded ranges indicate the location and FWHM of the RMSF at $\text{0}\,\text{rad}\,\text{m}^{-\text{2}}$ (i.e. the zero peak). \textsc{rm-clean} was applied to each phase-resolved FDF using a S/N threshold of 3; the \textsc{rm-clean} model components are shown as red dots.}
    \label{fig:phase-rm}
\end{figure*}

\subsubsection{Phase-resolved RM}
The apparent RM can vary significantly with pulse phase, which has been attributed to interstellar scattering, profile evolution with frequency, and magnetospheric effects \citep[e.g.][]{Noutsos2009,Noutsos2015,Dai2015,Ilie2019,Xu2025}.
We therefore measured the RM for the individual phase bins within the on-pulse region of each MSP.
We considered an RM measurement significant if the $1\sigma$ confidence interval (calculated via bootstrapping) was smaller than the FWHM of the RMSF, and the linear polarisation exceeded $5\sigma_I$ (where $\sigma_I$ is the standard deviation of the off-pulse noise).
To quantify the variation in RM, we calculated the maximum statistical difference between any two RM measurements over the profile.
The phase-resolved RM measurements and pulse profiles for the four MSPs with the most significant variations are shown in Figure~\ref{fig:pol-profs}.

Following the analysis by \citet{Noutsos2009}, we also measured the change in fractional circular polarisation over the observing band, $\Delta(V/I)$, by fitting a linear model $V/I\propto\nu$ to the spectral data for each phase bin.
Correlations between $\Delta(V/I)$ and $V$ as a function of pulse phase can be an indication of interstellar scattering \citep[see also][]{Ilie2019}.
Additionally, propagation through relativistic plasma in the pulsar magnetosphere can cause a frequency-dependent conversion from linear into circular polarisation, known as Faraday conversion \citep{Kennett1998}.
The effect of Faraday conversion on the position angle depends on the physical environment, and is therefore commonly modelled using a generalised RM (GRM); i.e.
\begin{equation}
    \psi = \psi_0 + \mathrm{GRM}\lambda^\alpha,
\end{equation}
where $\alpha$ is the wavelength dependence.
\citet{Lower2024} demonstrated that incorrectly modelled Faraday conversion (e.g. assuming $\psi\propto\lambda^2$) can produce apparent phase-dependent RM variations.
A detailed search for Faraday conversion \citep[as in][]{Lower2024} is beyond the scope of this work.
However, we note that a correlation between $\Delta(V/I)$ and the RM could be an indication of Faraday conversion.

We observe the largest apparent RM variations for PSR J2145$-$0750 (see Figure~\ref{fig:pol-profs}).
The apparent RM is stable at $\sim$\qty{-2.2}{\radian\per\m\squared} for most of the profile, but jumps to $\sim$\qty{0.5}{\radian\per\m\squared} near the centre of the trailing profile component.
The phase-resolved FDF (see Figure~\ref{fig:phase-rm}, right) shows that the polarisation is distributed between peaks at Faraday depths of $\sim$\qty{0.5}{\radian\per\m\squared} and $\sim$\qty{-2.2}{\radian\per\m\squared}.
It is particularly notable that between pulse longitudes of \qty{30}{\degree} and \qty{45}{\degree}, the polarisation is simultaneously present at both $\sim$\qty{0.5}{\radian\per\m\squared} and $\sim$\qty{-2.2}{\radian\per\m\squared} (as indicated by the \textsc{rm-clean} model components in Figure~\ref{fig:phase-rm}).
This strongly suggests that the peak closer to $\qty{0}{\radian\per\m\squared}$ is instrumental in origin.
To verify this hypothesis, we performed a jackknife cross-validation test by dividing the observation into four separate \qty{15}{\min} segments and re-performing the analysis on each segment.
As expected, the polarisation near $\sim$\qty{0.5}{\radian\per\m\squared} varies significantly in intensity over time, while the peak near $\sim$\qty{-2.2}{\radian\per\m\squared} remains stable.
We therefore conclude that the apparent variation is instrumental, and the true pulsar RM does not vary significantly with phase.

PSR J0437$-$4715 shows RM variations of $\sim$\qty{1}{\radian\per\m\squared} around the steepest part of the position angle curve (see Figure~\ref{fig:pol-profs}). In this case, the peak in the FDF appears to shift with phase (see Figure~\ref{fig:phase-rm}, left).
Additionally, the \textsc{rm-clean} model components indicate that -- unlike for J2145$-$0750 -- the FDF is accurately described by a single point source convolved with the RMSF at all phases.
This suggests that there is no significant achromatic instrumental polarisation.
As above, we performed a jackknife test by dividing the observation into two \qty{15}{\min} segments; in this case, we do not see any significant variations over time.
This is again consistent with there being no significant instrumental contamination in the signal.
Since this pulsar has a very low DM (\qty{2.64}{\per\cm\cubed\pc}), the estimated scattering timescale is much smaller than the bin width, which effectively rules out scattering as the cause of the variations.
Apparent RM variations can also be induced when the amplitudes of overlapping emission components evolve differently with frequency.
We investigated the profile evolution by averaging the frequency channels into three subbands and comparing the profiles; no significant profile shape variations were observed in the phase range of interest.
We therefore cannot attribute the variations to any known systematic effect.
We also do not measure any notable correlation between $\Delta(V/I)$ and the change in RM, so our analysis does not show any evidence of Faraday conversion.
Therefore, the origin of the variations is as-yet unclear.
Observing at lower frequencies or over a larger frequency range would help to rule out weakly-chromatic instrumental effects by reducing potential confusion in the FDF.
A larger fractional bandwidth would also improve the sensitivity to changes in $\Delta(V/I)$.
These requirements could be met by using non-contiguous frequency configurations \citep[e.g.][]{Kaur2022}.

Considering only the pulsars with unambiguous RM measurements (i.e. $\mathrm{RM}\gg \delta\phi/2$), we see notable variations for PSRs J2241$-$5236 and J2256$-$1024.
For both pulsars, the phase-dependent RM variations are correlated with the rate of change of the position angle.
For J2241$-$5236, there is also a clear correlation between $\Delta(V/I)$ and the rate of change of circular polarisation.
These results are consistent with interstellar scattering.
The estimated scattering timescales for these pulsars are $\sim\numrange{10}{20}\,\unit{\us}$ at \qty{154}{\MHz} \citep[given the DMs of 11.4 and 13.8\,\unit{\per\cm\cubed\pc};][]{Bhat2004}, compared to a bin width of \qty{3.125}{\us} (i.e. $\sim$\numrange{3}{6} bins).
This is consistent with the temporal scale of the RM variations observed in Figure~\ref{fig:pol-profs}.

It is interesting to note that PSRs J0437$-$4715, J2145$-$0750, and J2241$-$5236 all show larger RM variations in the Parkes data from \citet{Dai2015} than in the MWA data.
\citet{Noutsos2015} showed that smaller apparent RM variations at low frequencies are a natural result of interstellar scattering.
However, since both J0437$-$4715 and J2145$-$0750 have scattering timescales smaller than the bin width, this cannot explain the inconsistency with Parkes for these pulsars.


\subsection{Pulse profiles}

\subsubsection{Polarimetry}
For each of the detections, we generated full-polarimetric integrated pulse profiles.
We followed the methods described by \citet{Spiewak2022}, based on the methods from \citet{Everett2001}, to de-bias the linear polarisation measurements and to compute the uncertainties in the position angle measurements.
For detections with measured RMs, we removed the Faraday rotation using the \textsc{pam} utility in \textsc{psrchive}.

\begin{figure}[t]
    \centering
    \includegraphics[width=\linewidth]{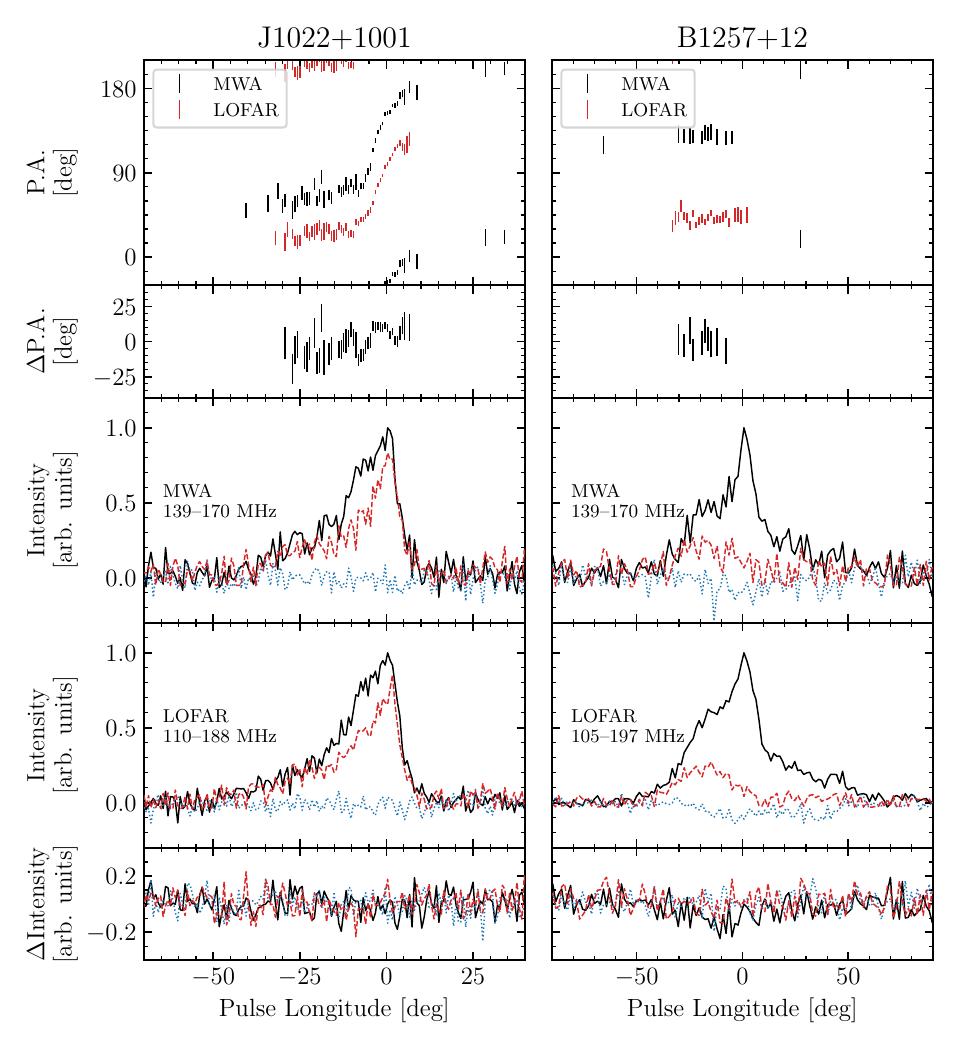}
    \caption{Comparison between MWA and LOFAR polarimetry for PSRs J1022+1001 (left) and B1257+12 (right). The LOFAR data are from \citet{Noutsos2015} and span a larger bandwidth, but have similar centre frequencies to the MWA data. The top two panels show the position angles (P.A.) and the residuals ($\Delta$P.A.) with the mean subtracted (the mean residual is arbitrary). The bottom three panels show the polarimetric integrated pulse profiles (the colours the same as in Figure~\ref{fig:pol-profs}) and the residuals in intensity.}
    \label{fig:compare-pol}
\end{figure}

In Figure~\ref{fig:compare-pol}, we show a comparison between the MWA and LOFAR polarimetric pulse profiles \citep[from][]{Noutsos2015} for PSRs J1022+1001 and B1257+12.
The profiles were aligned by minimising the standard deviation of the total intensity profile residuals.
As neither data were calibrated for absolute polarisation, the mean residual position angle is arbitrary.
For J1022+1001, there are negligible differences in the total and polarised intensities, but some systematic variations in the position angle residuals where the position angle is changing most rapidly with phase.
This may be due to frequency evolution of the position angle over the much larger LOFAR observing band.
For B1257+12, the position angles and polarised intensities are consistent, but there is a difference in the total intensity profiles at the leading edge which is likely due to frequency evolution of the pulse shape.
This is the first published verification of MWA-VCS polarimetry using an independent instrument at the same frequency.

\subsubsection{Frequency evolution}
\begin{figure*}[p]
    \centering
    \includegraphics[width=\linewidth]{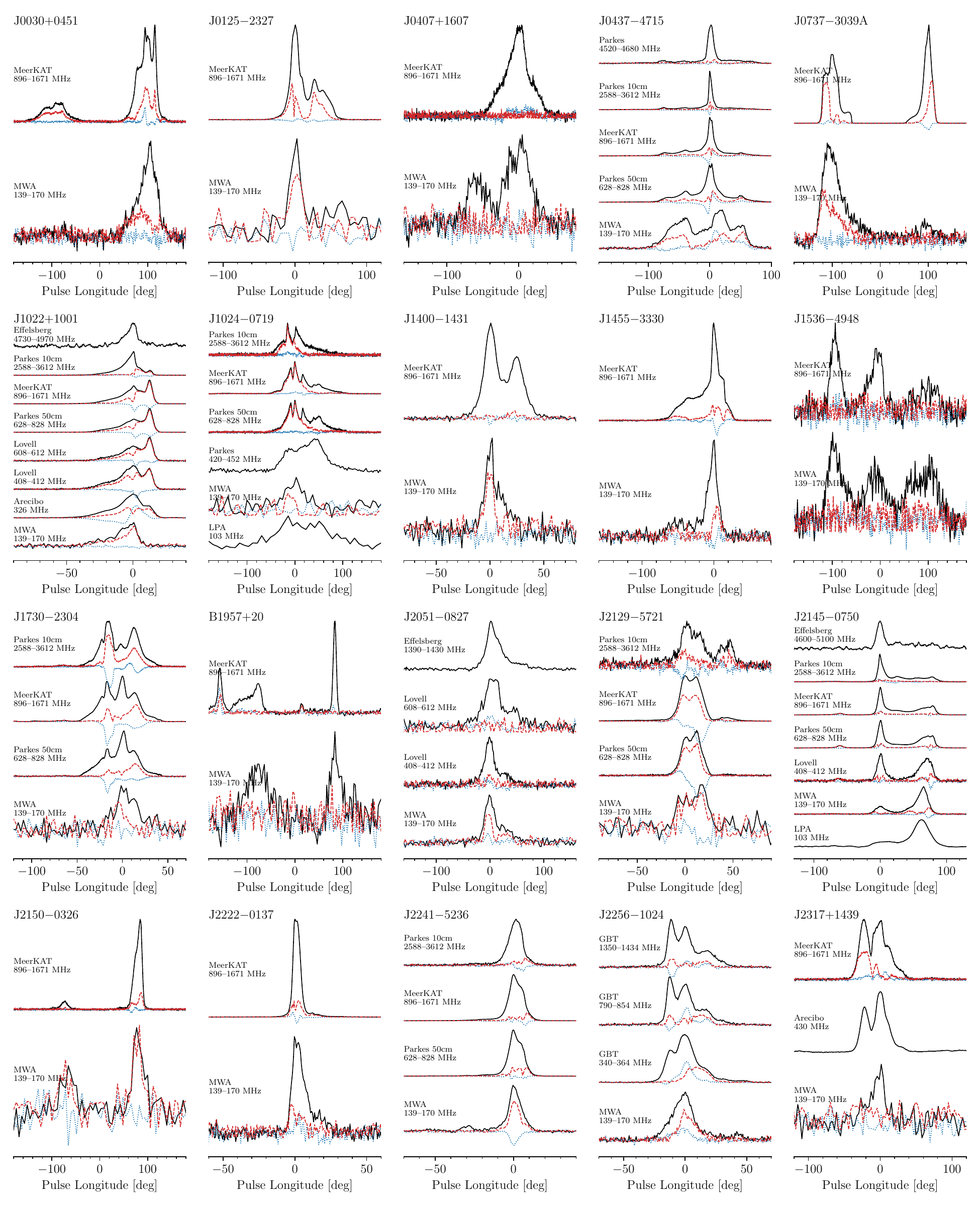}
    \caption{Frequency evolution of integrated pulse profiles for selected MSPs. MWA profiles are from this work. Profiles from other telescopes are from the following publications: MeerKAT -- \citet{Spiewak2022}; Murriyang/Parkes -- \citet{Manchester1995,Bailes1997,Dai2015}; Effelsberg -- \citet{Kijak1997,Kramer1998}; Lovell -- \citet{Stairs1999}; LPA -- \citet{Kuzmin1999}; GBT -- \citet{Crowter2020}; Arecibo -- \citet{Camilo1993,Wahl2023}.}
    \label{fig:stack-profs}
\end{figure*}

Characterising the frequency evolution of MSP pulse profiles is important for both pulsar timing and understanding the pulsar emission mechanism.
In this section, we provide a qualitative comparison of the MWA pulse profiles to archival profiles at other frequencies.
We compiled profiles from the MPTA census data release \citep{Spiewak2022} and the European Pulsar Network Database of Pulsar Profiles\footnote{\url{https://psrweb.jb.man.ac.uk/epndb/}}.
For PSR J2256$-$1024, we used the coherently-dedispersed GBT profiles from \citet{Crowter2020}.
Since timing information for the archival profiles was not available, the profiles were aligned by eye based on our own interpretation of the profile features, particularly using components common across multiple frequencies.
Figure~\ref{fig:stack-profs} shows pulse stacks for 20 selected MSPs; we particularly focus on those with interesting frequency evolutions and/or high-quality MWA profiles.
In the following, we describe the salient features and frequency evolution of the profiles for each of the selected MSPs.

\paragraph{PSR J0030+0451.}
The MWA profile suggests that both the leading and trailing components of the main pulse in the MeerKAT profile weaken at low frequencies.
This leads to a simpler, single-peaked main pulse.
The interpulse is also much weaker (though still detectable) in the MWA profile.

\paragraph{PSR J0125$-$2327.}
Although the \snr{} is low in the MWA profile, it is clear that the the trailing component is significantly weaker than in the MeerKAT profile.
The components also appear to be more separated in the MWA profile, leading to a larger profile width.
The MWA profile also shows a higher degree of linear polarisation.

\paragraph{PSR J0407+1607.}
The MWA profile shows a new leading component that is not seen in the MeerKAT profile.
Since profile component separations are known to change very slowly with frequency for MSPs, the two components in the MWA profile are not likely to be a bifurcation of the MeerKAT profile (as one may expect for a non-recycled pulsar).
Rather, this appears to be an entirely new component which has been enhanced at low frequencies.

\paragraph{PSR J0437$-$4715.}
The dramatic frequency evolution of this pulsar has previously been described by \citet{Bhat2018}.
The spacing and relative amplitude of the outer profile components evolve slowly with frequency, whilst the central peak appears to weaken and shift towards the trailing edge of the profile.
Interestingly, the inversion in circular polarisation near the centre of the profile which is observed at higher frequencies is not seen by the MWA.

\paragraph{PSR J0737$-$3039A.}
This is the MSP from the famous double pulsar system, which is a high-priority target for relativistic binary timing programs \citep[e.g.][]{Kramer2021MNRAS}.
This pulsar is significantly scattered at MWA frequencies, and as such a detailed characterisation of the scattering properties will be presented in a separate publication.
We note here that the relative amplitude of the main pulse and interpulse changes significantly between the MeerKAT and MWA profiles, implying different spectral indices.

\paragraph{PSR J1022+1001.}
This pulsar has a particularly interesting profile evolution in which the trailing component is prominent between $\sim$\qty{400}{\MHz}--\qty{2}{\GHz} and weakens at higher and lower frequencies.
We aligned the MWA profile with the leading component, consistent with the interpretation of \citet{Noutsos2015}.
This alignment implies that the trailing component is completely missing in the MWA frequency band.

\paragraph{PSR J1024$-$0719.}
The frequency evolution for this pulsar is perplexing.
There is a clear enhancement of the trailing component in the Parkes at \qty{436}{\MHz} compared to higher frequency profiles.
However, the MWA (and LOFAR) profile shows only a single component.
The LPA profile at \qty{103}{\MHz} then hints at two components (although it should be cautioned that this is an incoherently dedispersed profile which may be broadened due to intrachannel smearing).
There is therefore an ambiguity as to whether the component visible in the MWA profile corresponds to the leading or trailing component of the Parkes and LPA profiles.
The linear polarisation provides a clue: the MWA profile is linearly polarised towards the leading edge, similarly to the higher-frequency profiles.
This implies that the MWA profile should be aligned with the leading edge of the other profiles, contrary to what was assumed by \citet{Kondratiev2016}.
This alignment is more consistent with the LPA profile, as should be expected given the relatively small difference in frequency.
The overall trend is therefore similar to PSR J1022+1001: the trailing component is enhanced between $\sim$\qty{400}{\MHz}--\qty{2}{\GHz}, and weakens at higher and lower frequencies.
The difference is that for J1024$-$0719, the trailing component appears to be enhanced again at even lower frequencies.

\paragraph{PSR J1400$-$1431.}
The MeerKAT profile shows two distinct peaks, whereas the MWA profile shows only one.
It is unclear whether the MWA profile should be aligned with the leading or trailing peak of the MeerKAT profile, but we favour the leading peak due to the asymmetry in the MWA profile.
This alignment implies that the trailing component is not visible in the MWA profile.
The MWA profile is also strongly linearly polarised, whereas the MeerKAT profile is completely depolarised.

\paragraph{PSR J1455$-$3330.}
The trailing side of the profile (in particularly the most prominent component) is enhanced in the MWA profile.
However, the pulse width remains similar to the MeerKAT profile.

\paragraph{PSR J1536$-$4948.}
The profile for this pulsar has three pulses which together span nearly the entire rotation phase.
The MWA profile is noticeably scattered compared to the MeerKAT profile, and the trailing pulse is enhanced compared to the two other pulses.

\paragraph{PSR J1730$-$2304.}
The profile has three main components, with the central component having a steeper spectral index than the two outer components.
The leading component is significantly weakened in the MWA profile, and the trailing component is enhanced compared with the Parkes 50-cm profile.

\paragraph{PSR B1957+20.}
This is the first black widow MSP to be discovered.
The MWA profile is scattered, but it is clear that the main pulse and interpulse have different spectral indices, with the amplitudes being more similar in the MWA profile.

\paragraph{PSR J2051$-$0827.}
The MWA profile shape is similar to the Lovell profile at \qty{410}{\MHz}.
Notably, this pulsar exhibits rapid depolarisation with frequency: the MWA profile shows significant linear polarisation, whereas the profile is depolarised at \qty{410}{\MHz}.

\paragraph{PSR J2129$-$5721.}
The profile for this pulsar comprises a pulse with two components which is present across all frequency bands, as well as a post-cursor pulse which is only visible at high frequencies.
The MWA profile is slightly broader and more linearly polarised than the Parkes 50-cm profile, but is otherwise consistent.

\paragraph{PSR J2145$-$0750.}
The frequency evolution of this pulsar has been studied by several authors \citep[e.g.][]{Kramer1999,Kondratiev2016,Bhat2018,DilpreetThesis}.
At low frequencies, the trailing component is significantly enhanced and shifts towards the central component, and the precursor component disappears.
Both the linear and circular polarisation in the MWA profile is consistent with the Parkes and Lovell profiles.
The position angle curve (see Figure~\ref{fig:pol-profs}) shows an orthogonal polarisation mode jump (i.e. a jump of \qty{90}{\degree}), but is otherwise relatively flat.

\paragraph{PSR J2150$-$0326.}
The relative amplitude between the main pulse and interpulse changes considerably between the MeerKAT and MWA profiles, with the amplitudes being comparable in the MWA profile.
Again we observe rapid depolarisation, with both pulses in the MWA profile being completely linearly polarised, whilst the MeerKAT profile is only marginally polarised.

\paragraph{PSR J2222$-$0137.}
This is a mildly-recycled relativistic binary pulsar with a very low DM of \qty{3.27}{\per\cm\cubed\pc} \citep{Boyles2013}.
The MWA profile is broader than the MeerKAT profile, which is partly due to an enhancement of the post-cursor components.
The main component also appears wider in the MWA profile.

\paragraph{PSR J2241$-$5236.}
This pulsar has a very narrow profile which shows little variation with frequency.
The profile is weakly polarised at high frequencies, but is significantly polarised (both linearly and circularly) at MWA frequencies.
As first noted by \citet{Kaur2019}, the precursor components visible in the Parkes profiles with very low amplitude are quite prominent in the MWA profile and are offset earlier in phase.
Also notable is the trailing component visible as a shoulder in the MeerKAT and Parkes profiles, which cannot be seen in the MWA profile.

\paragraph{PSR J2256$-$1024.}
There is a clear enhancement in the central profile component at lower frequencies, to the extent that the leading and trailing components cannot be seen in the MWA profile.
We also see an enhancement in a post-cursor component in the MWA profile.
The post-cursor can be seen with a much lower amplitude in the \qty{352}{\MHz} GBT profile.

\paragraph{PSR J2317+1439.}
The leading profile component that is prominent at higher frequencies is significantly weaker in the MWA profile.
This can be seen more clearly in the LOFAR profile \citep[see][]{Kondratiev2016}.
\bigskip

In general, the profile component separations and widths appear to vary very little with frequency, whereas the relative amplitudes vary considerably.
This is different from non-recycled pulsars, which often show a separation of pulse components at low frequencies.
Our results are consistent with \citet{Kramer1999} and \citet{Kondratiev2016}, who attribute the slow frequency evolution of MSP profiles to the emission originating from a narrow altitude range due to the more compact magnetospheres.

A common trend observed for MSPs is asymmetric evolution of the profile components.
\citet{Dyks2010} suggested that enhancement of the trailing side of the profile at low frequencies can be attributed to aberration and retardation, whilst weakening of the trailing side is due to asymmetry in curvature radiation with respect to the dipole axis.
Therefore, the observed frequency evolution can indicate the dominant of these effects.
\citet{Kondratiev2016} identified a clear enhancement of the trailing side for PSRs J1022+1001, B1257+12, J2145$-$0750, and J2317+1439.
Our observations support these results; however, we also observed that the trailing side of PSR J1022+1001 weakens again at low frequencies (i.e. the component amplitude turns over).
In addition, we identify an enhancement of the trailing side of the profile for PSRs J1455$-$3330, J2222$-$0137, and J2256$-$1024.
We also identify weakening of the trailing side for PSRs J0030+0451, J0125$-$2327, J1400$-$1431, J2129$-$5721, and J2241$-$5236.
Overall, there are a comparable number of MSPs that exhibit each of the two trends.

We consistently observe that the MWA profiles are comparably or more polarised than higher-frequency profiles.
The depolarisation of pulsar emission with frequency is a well-known trend \citep[e.g.][]{Manchester1973,Kramer1999}, and is commonly attributed to the superposition of orthogonal polarisation modes \citep{vonHoensbroech1998,McKinnon1997}.
However, the specifics of the depolarisation mechanism are not agreed on, and it is likely that a combination of magnetospheric, co-rotational, and physical effects determine the polarisation evolution with frequency \citep{Noutsos2015}.
Regardless of the physical explanation, the observation of this trend provides an indirect verification of MWA-VCS polarimetry.
    \section{Limitations and Future Outlook}\label{sec:future}
The results of this census are close to the best that can be achieved with the current MWA-VCS.
In particular, all detections have been coherently beamformed and coherently dedispersed using voltage data from the nearest SMART observation to each pulsar, and calibrated with the latest-generation software and sky models.
Further improvements in \snr{} could be gained from targeted observations; however, due to the large data rate of the VCS (\qty{56}{\TB\per\h} for the Phase-II array) and the resources required to process VCS data, such observations are not practical, particularly for long-term monitoring.
A real-time beamformer is currently under development for the Phase-III MWA, which will provide more flexibility to perform targeted observations.
The Phase-III MWA also has simultaneous access to up to 256 tiles, which is double the collecting area of the Phase-II system.
These upgrades will in principle make pulsar monitoring tractable in terms of data volume, and the improved signal fidelity will make timing observations more scientifically useful.

The high DM precision obtainable from individual MWA-VCS observations is clearly demonstrated in Table~\ref{tab:detections}.
Even using a simple \snr{}-maximisation algorithm (as implemented in \textsc{pdmp}), we consistently obtain DM estimates with \qty{e-3}{\per\cm\cubed\pc} or better precision.
The prospects of further improving precision using template matching and more optimal frequency configurations are strong.
This has been demonstrated by \citet{Kaur2022}, who obtained precisions of up to \qty{e-6}{\per\cm\cubed\pc} for PSR J2241$-$5236 using MWA-VCS observations spanning \numrange{80}{220}\,\unit{\MHz}.
Even a precision of $\sim$\qty{e-4}{\per\cm\cubed\pc} is sufficient to monitor temporal variations in the DM due to the IISM and the solar wind, which is a requirement for high-precision pulsar timing \citep[e.g.][]{Tiburzi2021,Susarla2025}.
There are several timing-array MSPs that are either beyond the reach of LOFAR or may be better targets for the MWA due to their southern declinations.
For those that are not significantly scattered -- namely PSRs J0125$-$2327, J0437$-$4715, J1455$-$3330, J2039$-$3616, J2129$-$5721, and J2241$-$5236 -- the MWA is well positioned to improve current DM models.

Given that many of the detections presented here are the first at low frequencies, they will be valuable for measuring spectral indices and investigating low-frequency turnovers in flux density spectra.
Measuring scattering timescales for timing-array MSPs is also valuable, as these measurements can be used to predict the improvement in timing precision at higher frequencies \citep[e.g.][]{Kramer2021MNRAS}.

Lastly, accurate models of the detectable pulsar population at SKA-Low frequencies will be an important contribution from the MWA and SMART.
For example, \citet{Xue2017} used population synthesis informed by MWA detections to estimate the number of pulsars detectable with the SKA-Low to be $(9.4 \pm 1.3)\times 10^3$.
With the larger pulsar sample detected in SMART, and more accurate spectral models, these simulations can be improved, which will help to inform future pulsar surveys with the SKA-Low.
    \section{Summary and Conclusions}\label{sec:conclusions}
We have performed a census of MSPs in SMART survey observations at \qty{154}{\MHz}.
A total of 40 MSPs were detected, 11 of which are the first published detections below \qty{300}{\MHz}.
The results are summarised as follows:
\begin{itemize}
    \item We measured mean flux densities for all detections. Comparison with LOFAR measurements in the same frequency band, when possible, shows consistent agreement within a factor of two of the estimated uncertainties.
    \item Based on a close comparison of two MSPs, the MWA-VCS polarimetry is consistent with published polarimetry from LOFAR. This is the first published verification of MWA-VCS polarimetry with an independent instrument in the same frequency band.
    \item We measured significant profile-averaged RMs for 25 MSPs. The measurements are consistent with MeerKAT and Parkes, with comparable or better precision.
    \item Three pulsars show apparent phase-dependent RM variations. We attribute the variations for PSRs J2241$-$5236 and J2256$-$1024 to interstellar scattering. For PSR J0437$-$4715, the variations appear to be intrinsic, but observations over a larger fractional bandwidth would help to confirm this.
    \item The MWA profiles provide further evidence that MSPs show remarkably little frequency dependence in the spacing of profile components but large changes in the relative component amplitudes. This is consistent with the emission originating from compact magnetospheres.
    \item We observe several clear examples of asymmetric profile evolution, which may be related to the two opposing co-rotational effects described by \citet{Dyks2010}.
    \item In general, the MWA profiles are similarly or more polarised than higher-frequency profiles. This is consistent with the expected trend, and provides an indirect verification of MWA-VCS polarimetry.
\end{itemize}
Many of the MSPs detected in this census are high-priority targets for pulsar timing array experiments.
Future upgrades in MWA Phase III are expected to enable regular monitoring of these targets for the first time.
This will be an important step to achieve as we move towards the era of the SKA-Low.
    {%
\small

\paragraph{\bfseries Acknowledgements.}
We thank K.~Crowter for providing the GBT pulse profiles for PSR J2256$-$1024.
We also thank the anonymous reviewer for insightful comments.
C.P.L. was supported by an Australian Government Research Training Program (RTP) Stipend and RTP Fee-Offset Scholarship.
This scientific work made use of Inyarrimanha Ilgari Bundara, the Murchison Radio-astronomy Observatory, operated by CSIRO.
We acknowledge the Wajarri Yamatji people as the traditional owners of the Observatory site.
This work was supported by resources provided by the Pawsey Supercomputing Research Centre with funding from the Australian Government and the Government of Western Australia.
This work made extensive use of the ATNF Pulsar Catalogue \citep{Manchester2005}, the European Pulsar Network Database of Pulsar Profiles, NASA's Astrophysics Data System, and arXiv.

Software/packages:
\textsc{dspsr} \citep{vanStraten2011},
\textsc{psrchive} \citep{Hotan2004,vanStraten2012},
\textsc{spinifex} \citep{spinifex},
\textsc{matplotlib} \citep{Matplotlib},
\textsc{numpy} \citep{NumPy},
\textsc{scipy} \citep{2020SciPy-NMeth},
\textsc{astropy} \citep{Astropy2013,Astropy2018,Astropy2022},
\textsc{psrqpy} \citep{psrqpy}.

\paragraph{\bfseries Data Availability.}
The coherently-dedispersed data cube archives for all 40 detected MSPs are available in \textsc{psrchive} format with the frequency and time resolutions described in Section~\ref{sec:processing}.
The polarisation profiles are also available in \textsc{ascii} format.
We have also provided the data from Tables~\ref{tab:detections} and \ref{tab:rm} in a machine-readable format.
The dataset can be accessed under the DOI 10.5281/zenodo.16628260.

}
    \printbibliography
\end{document}